\documentclass[sigconf]{acmart}
\usepackage{multirow}

\AtBeginDocument{%
  }

\copyrightyear{2026}
\acmYear{2026}
\setcopyright{cc}
\setcctype{by}
\acmConference[CHI '26]{Proceedings of the 2026 CHI Conference on Human Factors in Computing Systems}{April 13--17, 2026}{Barcelona, Spain}
\acmBooktitle{Proceedings of the 2026 CHI Conference on Human Factors in Computing Systems (CHI '26), April 13--17, 2026, Barcelona, Spain}
\acmPrice{}
\acmDOI{10.1145/3772318.3790557}
\acmISBN{979-8-4007-2278-3/2026/04}

\begin{document}

\title{Finger Tendon Vibration: Finger Movement Illusions for Immersive Virtual Object Interaction}
\author{Kun-Woo Song}
\email{kwsong0725@kaist.ac.kr}
\orcid{0009-0003-9306-546X}
\affiliation{%
  \department{Graduate School of Culture Technology}
  \institution{KAIST}
  \city{Daejeon}
  \country{Korea}
}

\author{Youngrae Kim}
\email{dufo7070@kaist.ac.kr}
\orcid{0009-0005-7363-6047}
\affiliation{%
  \department{Graduate School of Metaverse}
  \institution{KAIST}
  \city{Daejeon}
  \country{Korea}
}

\author{Sang Ho Yoon}
\email{sangho@kaist.ac.kr}
\orcid{0000-0002-3780-5350}
\affiliation{%
  \department{Graduate School of Culture Technology}
  \institution{KAIST}
  \city{Daejeon}
  \country{Korea}
}

\renewcommand{\shortauthors}{Song et al.}

\begin{abstract}
    The absence of physical information during hand-object interaction in a virtual environment diminishes realism and immersion. Kinesthetic haptic feedback has proven effective in delivering realistic object-derived haptic cues, enhancing the overall virtual reality~(VR) experience. Here, we propose kinesthetic illusion through a novel application of finger tendon vibration~(FTV), which creates an illusory sensation of finger movement. To effectively apply FTV for virtual object interactions, we first examine the effects of short-duration FTV (<5~s) through 3~perception studies. Based on study results, we design 6~exemplary VR scenarios, representing the overall design space of VR object interactions, and 4~different haptic rendering strategies for FTV. We evaluated these rendering methods on each VR scenario and derived a design guideline for FTV application. We then compared FTV with no vibration and simple vibration, observing that FTV enhances VR experience by providing realistic resistance on the finger, greatly improving body ownership. 
\end{abstract}


\begin{teaserfigure}
  \includegraphics[width=\textwidth]{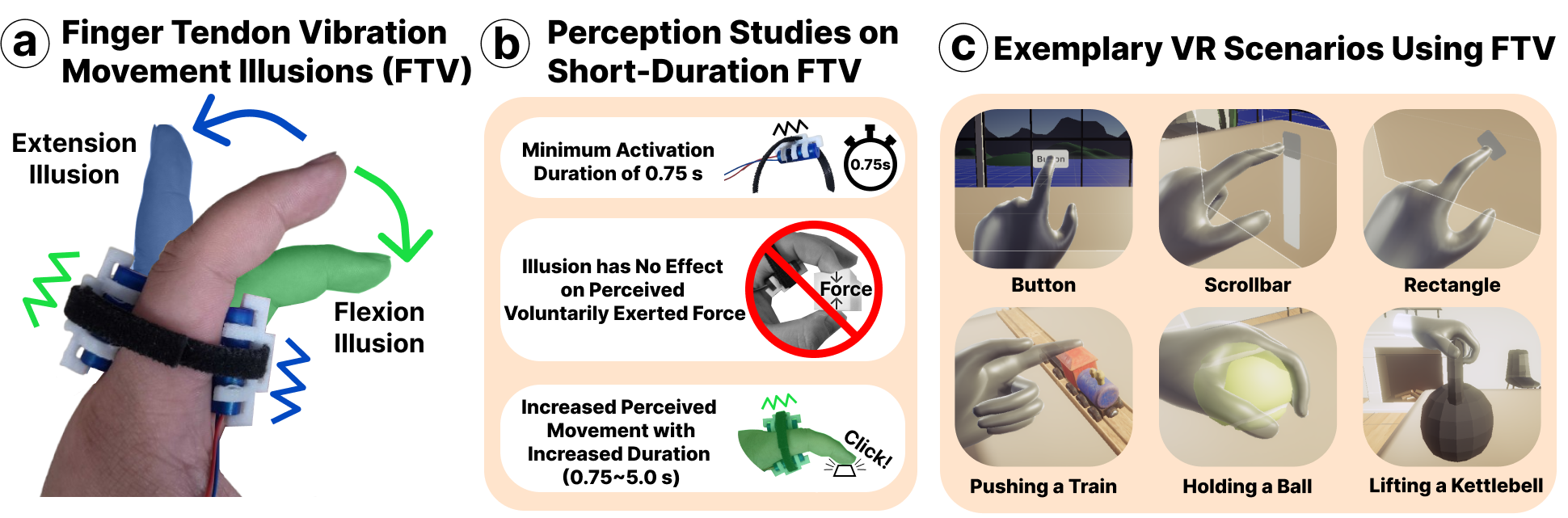}
  \caption{Overview of finger tendon vibration~(FTV). (a)~We propose an illusory sensation of finger extension or flexion without physically moving the finger. (b)~We conducted 3~perception studies on short durations~($<$~5~s) of FTV. We found that FTV requires a minimum vibration duration of 0.75~s to produce a perceptible illusion, short-duration FTV has no effect on the perceived voluntarily exerted force, and the illusory movement by FTV increases with increased duration. (c)~We designed 6~virtual object interaction scenarios utilizing FTV for immersive VR experiences.}
  \label{fig:teaser}
\end{teaserfigure}

\maketitle

\section{Introduction}~\label{sec:intro}
As virtual reality~(VR) systems evolve, the demand for natural and realistic interaction with their content constantly grows. 
A key interaction involves virtual objects, as users interact with them in manners that replicate real-world touch and manipulation.
By providing realistic tactile cues, kinesthetic feedback enhances the overall VR user experience by improving realism, immersion, and presence~\cite{sallnas2000supporting, gibbs2022comparison}. 
Kinesthetic haptic devices steadily evolved from grounded devices~\cite{massie1994phantom} to more recently mobile devices such as handheld~\cite{zenner2017shifty} and wearables~\cite{zhou2014rml}. 

However, to provide sufficient mechanical force for realistic feedback, these devices tend to become complex with intricate components such as cables~\cite{zhou2014rml} and air propellers~\cite{heo2018thor}. 
This makes it challenging to integrate with pre-existing hardware, and their bulky size causes fatigue in users during prolonged use.
To circumvent disadvantages regarding hardware form factors, researchers focused on building devices with smaller sizes and simple hardware setups by providing kinesthetic haptic illusions such as pseudo-haptics~\cite{samad2019pseudo}, skin stretch~\cite{kim2024quadstretcher}, grain vibration~\cite{heo2016vibrotactile}, and asymmetric vibration~\cite{tappeiner2009good}. 
Likewise, we employ tendon vibration-induced finger movement illusion with simple hardware configuration~(2 vibration motors).

We propose a novel use of finger tendon vibration~(FTV) to provide realistic haptic cues in VR for virtual object interactions~(Figure~\ref{fig:teaser}a). 
Muscle and tendon vibration stimulates muscle spindle endings, which detect muscle stretch. 
Depending on the muscle location, the vibro-stimulation results in illusory kinesthetic movements and force sensations without physical movement~\cite{goodwin1972proprioceptive, gilhodes1986perceptual, jones1985effect}. 
Researchers previously have conducted perception studies using vibrations of long durations~\cite{taylor2017muscle}. 
Researchers have recently begun applying muscle and tendon vibrations as a haptic feedback method. 
In VR, larger muscles and tendons on limbs have been used to enhance pseudo-haptic sensations~\cite{hirao2023leveraging} and create ground sway illusions~\cite{narita2024kinesway}.
However, FTV has been limitedly explored on the finger and has been applied only for hand rehabilitation, which used FTV-induced movement illusions for mirror therapy~\cite{gay2007proprioceptive, rinderknecht2013combined}. 
To the best of our knowledge, there have been no cases of employing FTV as a mode of haptic feedback.
There is still a lack of information on short-duration FTV, such as the minimum required duration for perceiving FTV and whether short-duration FTV is long enough to effectively provide distinguishable illusion sizes.

In this paper, we apply FTV to the index finger to induce illusory flexion and extension movement.
To enable FTV as haptic feedback for virtual object interactions, we explored how people perceive short durations of FTV through 3~perception studies on the required minimum duration to perceive FTV, the effect on perceived voluntarily exerted force, and the effect on perceived involuntary finger movement. 
From the 3~studies, we found: FTV requires a minimum duration of 0.75~s; short-duration FTV does not have a significant effect on the perceived voluntarily exerted force; the illusory movement caused by FTV increases over time; and 0.75~s of FTV causes an illusory movement similar to that of pressing a key on a keyboard~(Figure~\ref{fig:teaser}b).
Based on study results, we formulated 6~different VR scenarios exemplifying key virtual object interaction characteristics, including user interface~(UI) components such as pushing a button, moving a scrollbar, and dragging a 2D~rectangle and 3D objects such as pushing a train, holding a ball, and lifting a kettlebell~(Figure~\ref{fig:teaser}c).
For these 6~scenarios, we evaluated different haptic rendering methods based on the perception study results, including 2~different FTV triggering methods, \textit{Collision} and \textit{Predict}, and 2~FTV duration settings, \textit{Full} and \textit{Fixed}. 
From the rendering evaluation, we found: for sufficiently long interactions, the delay in illusion onset is negligible; and the tactile feedback from prolonged vibrations, despite the extra vibration having no impact on the perceived illusion, is essential for situations with consistent force.
Using the most effective rendering methods, \textit{Full} and \textit{Collision}, we compared FTV with simple vibration and no vibration to observe how well FTV improves VR experience. 
We observed that FTV significantly improves VR experience, particularly for body ownership by providing a feeling of resistance with direction.
Lastly, using the observations from both the rendering method evaluation and VR experience evaluation, we create an interaction guideline for applying FTV to VR object interaction.

Our contributions are the following:
\begin{itemize}
    \item A novel idea of using short-duration FTV to implement kinesthetic haptic cues for virtual object interactions with associated motion illusions
    \item Findings from perception studies that assess whether short-duration FTV of less than 5~seconds provides sufficient illusion for real-time haptic feedback during immersive virtual object interaction
    \item Example use case scenarios demonstrating the use of FTV-induced finger motion illusions to promote realistic and immersive haptic experience for virtual object interactions with a practical interaction guideline for FTV haptic rendering methods 
    \item An evaluation for VR object interactions, showing that FTV is an effective method to increase overall VR experience. 
\end{itemize}

\section{Related Works}~\label{sec:relatedworks} 
\subsection{Haptic Feedback for Virtual Object Interactions} 
Kinesthetic haptic feedback has been broadly studied to enrich virtual object interactions and has shown improved realism, immersion, presence, and performance~\cite{sallnas2000supporting, gibbs2022comparison}. 
Realistic haptic feedback that matches the user's expectation based on real-life experience also increases body ownership~\cite{ferri2013body}.
Consequently, kinesthetic feedback has shown its effectiveness in virtual applications requiring information on the virtual object's physical properties, such as bone surgery training~\cite{gani2022impact} and virtual collaboration involving moving a virtual object together~\cite{basdogan2000experimental}.

Unlike vibrotactile feedback, which only affects the skin surface, kinesthetic feedback also engages muscles and joints. 
This deeper engagement has led to the development of different form factors, methods, and devices to provide such feedback realistically. 
Grounded devices such as PHANToM~\cite{massie1994phantom} and SPIDAR~\cite{sato2002development} simulated the stiffness and resistance of the virtual object with a high degree of realism. 
On the other hand, handheld and wearable devices apply mechanical methods that don't require as much space, such as weight shifts~\cite{zenner2017shifty} and air propulsion~\cite{heo2018thor}. 
Haptic gloves such as HaptX gloves~\cite{haptX}, which use pneumatic actuators, and RML Glove~\cite{zhou2014rml}, which use cables, mechanically interact with users' hands to deliver force haptic feedback. 
Electrical methods such as electrical muscle stimulation~(EMS)~\cite{lopes2017providing} and electrostatic~\cite{hinchet2018dextres}'s compact form serve as effective methods for wearable form factors.

Still, previous approaches required either bulky setups, which limited user mobility and introduced discomfort and fatigue over extended use, or complicated configurations, which often created difficulty when integrating with existing hardware.
To overcome fundamental weaknesses from physical hardware, we propose employing motion illusions from FTV to provide kinesthetic feedback with minimal hardware setup.

\subsection{Kinesthetic Haptic Illusions}
To tackle the weaknesses of kinesthetic haptic devices, previous researchers explored using a minimal amount of actuators to provide realistic haptic feedback by pushing the limits of human perception. 
Here, we provide an overview of methods involving no physical devices to smaller devices that deliver kinesthetic haptic illusions.

Some widely used methods include pseudo-haptics, which provide haptic feedback without the stimuli through other senses~\cite{pusch2011pseudo}. 
By creating visual offset, pseudo-haptics with physical devices could replace proprioceptive information with visual feedback~\cite{lecuyer2000pseudo}. 
More recent approaches involve only visual cues by manipulating control-display ratios evoking weight illusions~\cite{samad2019pseudo} and kinesthetic feedback using muscle tension as input~\cite{rietzler2019virtual}. 
For another method using physical stimuli, previous researchers utilized skin stretch on the forearm to provide movement~\cite{lin2024armdeformation} and force illusions~\cite{kim2024quadstretcher}. 
On a smaller scale, skin stretch on the finger showed how people perceived their fingers bending~\cite{collins2005cutaneous}.

Vibrations alone can also provide kinesthetic illusions. Through grain vibrations, researchers provided compliance feedback similar to pressing a button~\cite{heo2016vibrotactile} and objects twisting or bending~\cite{heo2019pseudobend}. 
Recent work further applied grain vibrations on base materials with different compliance to provide a wider variety of material force compliance~\cite{mun2025diversifying}.
Previous researchers also employed asymmetric vibration to provide force illusions~\cite{tappeiner2009good,tanabe2017evaluation} which further expanded to rotational movement~\cite{culbertson2017waves}, pulling illusions~\cite{tanabe2020effects}, and gravitational and inertial forces~\cite{choi2017grabity}.

In this paper, we employ a vibration-induced illusion that stimulates finger tendons to induce movement illusions. 
By using vibration motors, we minimize the size and weight of the haptic device and maintain simplicity while providing realistic haptic cues. 
Previously, tendon vibration on the arm has been used in VR to create kinesthetic illusions~\cite{hagimori2019combining} and to manipulate weight perception with visual feedback~\cite{hirao2023leveraging}. 
Still, to the best of our knowledge, no previous works have used tendon vibration on the finger as a means of haptic feedback to enhance immersion for virtual object interactions. 

\subsection{Finger Tendon Vibration}
Muscle spindles detect the length of the muscle and changes in muscle length. 
Early works discovered that we could ``trick'' these spindles to perceive a change in muscle length without any physical changes via vibrations~\cite{goodwin1972proprioceptive, gilhodes1986perceptual}. 
By vibrating muscle or tendon, the perception of a lengthened muscle creates illusory movement sensations~\cite{roll2009inducing} as well as illusory force sensations~\cite{jones1985effect}. 

To elicit this illusion, several vibration parameters come into play. 
First, the applied vibration frequencies between 70$\sim$110~Hz have proven to be effective at producing the illusion~\cite{schofield2015characterizing}. 
Also, vibrations with displacement amplitudes from 0.25~mm~\cite{albert2006ia} to 2~mm~\cite{craske1977perception} were enough to evoke the illusion. 
When applied to the wrist, perceived hand displacement and velocity increased as vibration durations increased~\cite{kito2006sensory, fuentes2012temporal}. 
Longer vibrations of 6~minutes caused a lasting aftereffect and a reverse illusion after the vibration ended~\cite{seizova2007proprioceptive}, but for shorter durations such as 15~s and 30~s, past works reported no aftereffects~\cite{tidoni2015illusory}.
In our work, we chose to fix the vibration frequency~(80~Hz) and amplitude~(1~mm) while manipulating the duration~(0.75$\sim$5.0~s) to enable the FTV illusion. 

For VR application, tendon vibration-induced movement and force illusions effectively enhanced kinesthetic illusion~\cite{hagimori2019combining} and pseudo haptic sensations~\cite{hirao2023leveraging}. 
For instance, the illusion helped simulate ground sway and tilt~\cite{narita2024kinesway} when applied to the ankle, and likewise, illusory movements on the neck muscles decreased VR sickness~\cite{song2025neck}.
More particularly, recently, there has been a greater focus on VR applications using arm tendon vibrations~\cite{cibulskis2025tendon, ushiyama2025kinesthetic}.

While these illusions focused extensively on larger muscles, little attention has been given to the finger. 
Collins et al. observed how vibrations of $2\sim4$~s on the dorsal side of the finger caused finger flexion sensation~\cite{collins2005cutaneous}. 
However, they did not report how the different durations, including shorter ones, affect the perception. 
FTV has been applied to hand rehabilitation by providing finger mobility sensation for mirror therapy~\cite{gay2007proprioceptive, rinderknecht2012device, rinderknecht2013combined}. 
Notably, Rinderknecht et al. displayed a virtual hand in front of participants that matched an extension movement illusion~\cite{rinderknecht2013combined}. 
Previous works on FTV agree that vibration on the palmar side of the proximal phalanx elicits a finger extension movement illusion on the metacarpophalangeal~(MCP) joint, while the dorsal side elicits a finger flexion movement illusion on the MCP joint~\cite{gay2007proprioceptive, rinderknecht2013combined}.
However, these works lack details for FTV that are required for VR applications.
First, the durations used in previous works were too long for short and sporadic VR hand interactions. 
Second, previous FTV research largely focuses on its illusion of involuntary finger movement, whereas for other body parts, it has been shown that tendon vibration also affects voluntarily exerted force~\cite{reschechtko2018force}.
Lastly, unlike arm or wrist tendon vibrations, FTV affects the whole finger, which consists of 3 phalanges, and previous literature does not specify how FTV affects each phalanx beyond the MCP joint.

To this end, we extend the use of FTV from hand rehabilitation to immersive virtual object interactions by promoting real-time haptic feedback. 
We conducted 3~perception studies on FTV focused on vibration durations shorter than 5.0~s. 
First, we investigated the minimum required vibration duration to elicit the illusion of finger flexion and extension movement. 
Then, we analyzed FTV's effects on participants' perceived voluntarily exerted force and involuntary finger movement to confirm whether short-duration FTV still provides illusions similar to those of longer tendon vibrations, along with how the illusion affects individual phalanges. 
We use these results to develop design guidelines for the use of FTV.

\section{Finger Tendon Vibration Principle}~\label{sec:principle}
In this section, we explain how we developed FTV to create a finger flexion and extension sensation. 
When applying tendon vibration on the biceps brachii, researchers observed that people perceived a greater illusion on the non-dominant arm~\cite{tidoni2015illusory}. 
To observe a greater illusion for short-duration FTV, we recruited only right-handed participants and applied FTV to their left index finger. 

\begin{figure}[t]
  \centering
  \includegraphics[width=\linewidth]{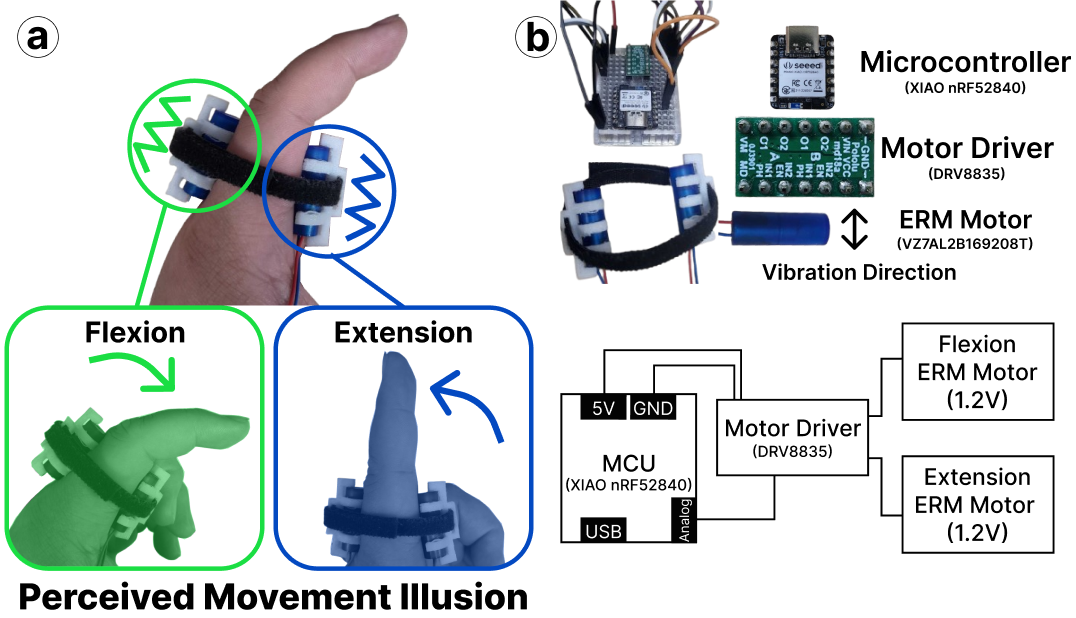}
  \caption{Overall system configuration. (a)~We placed vibration motors on the proximal phalanx of the left index finger. Vibrating the dorsal side of the finger results in a perceived flexion movement illusion, while the palmar side results in a perceived extension movement. (b)~We used a microcontroller, motor driver, and 2~eccentric rotating mass motors to form FTV.}
  \label{fig:implementation}
\end{figure}

\subsection{Implementation}
\textbf{Actuator Configuration.} 
Referring to previous papers using FTV~\cite{rinderknecht2013combined, gay2007proprioceptive}, we chose the proximal phalanx of the left index finger as the vibration site~(Figure~\ref{fig:implementation}a). 
To induce a flexion movement, we placed the vibration motor on the palmar side of the proximal phalanx. 
For an extension movement, we placed the motor on the dorsal side near the knuckle. 
In our work, to unify the terms, we refer to these motors, their locations, and the resulting illusion with the respective illusory movement directions~(flexion or extension). 

\textbf{Hardware.} 
To reliably produce movement illusions by tendon vibrations, the actuator needs to deliver perpendicular force against the finger tendons at a frequency between 70$\sim$120~Hz and vibration displacement amplitude between 0.2$\sim$3.0~mm~\cite{taylor2017muscle,schofield2015characterizing}.
Furthermore, on the finger, the actuator needs to be small enough to rest on a single finger without touching neighboring fingers. 
If the actuator touches other fingers, the additional tactile information on the relative position of the finger and actuator would compromise the illusion, requiring more time to sense the illusion~\cite{lackner1984influence}.
As such, we employed two eccentric rotating mass~(ERM) motors~(VZ7AL2B169208T, Vybronics, 8.8~mm$\times$8.8~mm$\times$24.9~mm) and operated them at 80~Hz with a $\pm$1~mm vibration displacement amplitude. 

However, ERM motors inherently have a small delay time because of the inertia of the rotating mass.
Other types of motors, such as linear resonant actuators~(LRA), have faster response time, but we could not find one that could deliver the required forces while maintaining a small size.
To confirm the motor-induced delay is small enough for short-duration FTV, we measured the delay by recording the vibration onset with an accelerometer~(ADXL335, Analog Devices) for 0.5~s, 0.75~s, 1.5~s, 3.0~s, and 5.0~s vibration durations used in our perception studies.
We repeated the vibration 10 times for each duration and obtained an average start and end delay of 50.5~ms, which is well within the perceptual threshold for 90\% of the population to notice visual-haptic asynchrony~\cite{di2019perceptual}.
For the full results, refer to Appendix~\ref{appendix:erm}.

To prevent the vibrations from spreading to different locations, we placed the motors in a 3D-printed~(TPU 95A) motor housing~(Figure~\ref{fig:implementation}a).
To easily adjust different finger sizes, the motor housings were strapped to their locations using Velcro straps.
The total weight of the motor, motor housings, and straps was 13~g. 
We drove the ERM motors using a motor driver~(DRV8835, Texas Instrument) and a microcontroller~(XIAO nRF52840, Seeed Studio) to communicate vibration commands with the computer~(Figure~\ref{fig:implementation}b).

\textbf{Calibration.} During our initial test of FTV with different users, we observed that the locations where people perceived the greatest illusion were slightly different. 
Some would feel a greater movement when shifting the motor location slightly to the side or moving it higher or lower along the finger. 
As such, for our calibration process, we used a similar approach as previous work~\cite{tidoni2015illusory} where participants self-reported when they felt the illusion. 
We applied 5~random vibration durations~(1.5~s, 2.0~s, 2.5~s, and 3.0~s) for either flexion or extension. 
After each vibration, we asked participants to describe what movement or force they perceived and did not correct them so as not to bias their answers. 
If the location of the motor was not correct, the participant would not feel the movement illusion and would feel only vibration. 
When the participants failed to describe the expected direction or the illusory effects of the FTV, we changed the motor location and repeated with 5~different random vibrations until they answered all 5~directions correctly. 
Equipping the device with calibration typically took less than 10 minutes.

\subsection{Preliminary Perception Study: Minimum Duration Threshold for FTV}
Before exploring how people perceive FTV illusions, we conducted a preliminary perception study to find the minimum vibration duration threshold for creating the illusion. 
We conducted a double random staircase experiment to obtain the FTV duration threshold for flexion and extension illusions.

\begin{figure}[t]
  \centering
  \includegraphics[width=\linewidth]{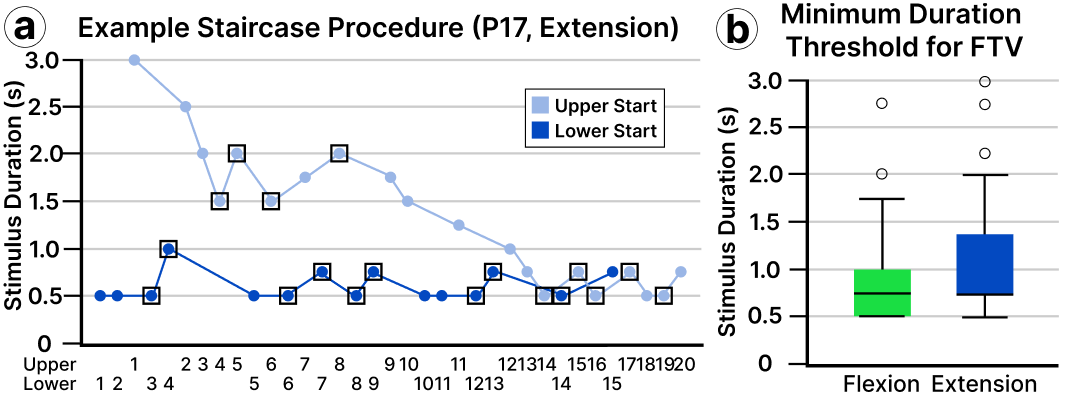}
  \caption{(a)~An example of double random staircase procedure~(P17, extension). Black squares mark reversals. (b)~Boxplot of final stimulus duration for the extension and flexion directions. Circles denote outliers.}
  \label{fig:staircase}
\end{figure}

\subsubsection{Procedure}
We recruited 31 participants~(15 male, 16 female; average age: 26.1, SD: 2.58). 
To detect the minimum duration threshold, we vibrated the motor for a target duration and asked the participants to report whether or not they felt the FTV illusion. 
Our staircase was a simple up-down with a minimum value of 0.5~s and a maximum of 3.0~s. The step size was 0.5~s and decreased to 0.25~s after an initial 3~reversals. 
There was a 5-second rest between each response and the next stimulus. After 6~reversals for the decreased step size, the staircase concluded~\cite{levitt1971transformed}. 
We also terminated the staircase if the response did not change at the extreme values~(0.5~s and 3.0~s) for 6~times in a row.
To prevent bias from a single staircase, we intertwined the two staircases in random order. 
We used an upper staircase, which started at the maximum value of 3.0~s, and a lower staircase, which started at 0.5~s, and proceeded either staircase randomly~(Figure~\ref{fig:staircase}a). 
Each participant completed one set of double random staircases for each illusion direction.
Odd-numbered participants completed the extension staircases first, and even-numbered participants completed the flexion staircases first. 
After completing the first direction, participants took a 3~minute break before moving on to the other direction.
Each participant took up to 15~minutes to complete the study.

\subsubsection{Results}
On average, the extension set of staircases took 45.9 trials to complete, and the flexion set of staircases took 45.3 trials to complete. 
No participants terminated the staircase by not sensing any difference at the extreme point for 6~responses.
The final stimulus duration for each staircase for each participant is shown in Figure~\ref{fig:staircase}b. 
Due to the skewed distribution of the final stimulus duration, we used the median for each staircase, which was 0.75~s for both directions. 
We set this value as our minimum threshold estimate for future use.

\begin{figure}[t]
  \centering
  \includegraphics[width=\linewidth]{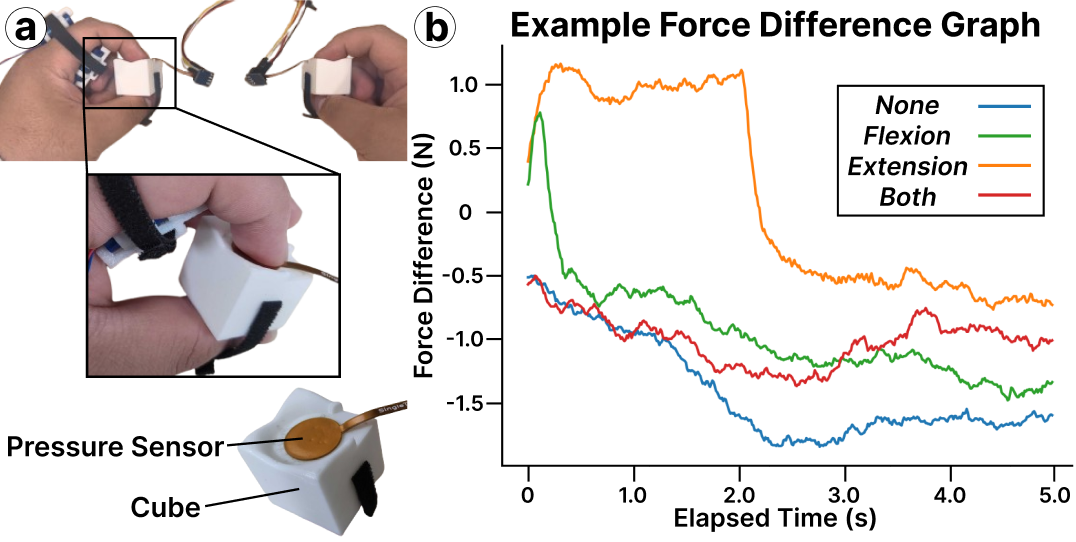}
  \caption{Perception Study~1 setup and results. (a)~To measure the perceived index pinching force, participants held a $2.5$~cm$\times2.5$~cm$\times2.5$~cm cube with the pressure sensor between their index finger and the side of their thumb. (b)~Example force difference~(P13) between the right hand~(matching hand) and the left hand~(FTV).}
  \label{fig:pinch}
\end{figure}

\section{Finger Tendon Vibration Perception Studies}
Tendon vibration in other areas, such as the arm, showed a significant effect on force~\cite{reschechtko2018force} and movement~\cite{hirao2023leveraging}. 
FTV has also shown a sufficient range of movement to be effective in rehabilitation~\cite{gay2007proprioceptive, rinderknecht2013combined}. 
However, vibration durations were too long for typical VR object interactions, spanning from 15~s~\cite{rinderknecht2013combined} to 1~min~\cite{gay2007proprioceptive}. 
To observe the real-time effects of FTV under shorter vibration durations~($<$5.0~s), we conduct perception studies on how short-duration FTV affects user perception of voluntarily exerted force, through pinching force, and involuntary movement. 

We recruited 28 participants~(13 male, 15 female; average age: 27.9, SD: 5.16) for the 2 perception studies, which took less than an hour. 
Initially, calibration took about 10 minutes. 
Then, the first perception study took 20 minutes and participants took a 5-minute break between the first and second perception study. 
The second perception study took another 20 minutes.

\subsection{Perception Study 1: Effect on Perceived Voluntarily Exerted Force}
In the first perception study, we measured how FTV affects people's perception on their voluntarily exerted force through the user's pinching force.
We adopt a similar matching test conducted by Reschechtko et al.~\cite{reschechtko2018force} where participants pushed on a force plate on a table, where they pushed on the plate with their task hand with a target force and tried to match the exerted force simultaneously on the opposite matching hand while their arm was vibrated on the extensor muscles. 

Participants held a $2.5$~cm$\times2.5$~cm$\times2.5$~cm cube with an attached pressure sensor~(S15-4.5N, SingleTact, 100Hz) in their left and right hands between the thumb and the index finger to measure the exerted force~(Figure~\ref{fig:pinch}a). 
Participants sat in front of a monitor that displayed the raw force data for training sessions.

Reschechtko et al.~\cite{reschechtko2018force} observed that when vibrating the task hand, which causes a flexion movement illusion, the opposite matching hand overestimated and exerted more force.
Likewise, if FTV has an effect on the exerted force, delivering FTV in different directions would alter the perceived force in the matching finger. 
If the results on the arm can be generalized to finger tendons, we can expect a flexion FTV on the task hand would cause the matching hand to exert more force when pinching.
For comparison, we also observed effects of no vibrations~(\textit{None}) and vibrations on both sides~(\textit{Both}).

\subsubsection{Procedure}
In our study, we fixed the left index finger as the task finger and the right index finger as the matching finger.
The study was a repetition of a training session and a matching session. 
In the training session, we displayed the raw force data of the left index force sensor so the participants confirmed how much pinching force they were exerting. 
We tasked the participants to memorize the amount of pinching force required for 5.0~N. 
Participants practiced the target pinching force for 30~s$\sim$90~s until they were confident to move on.

In the matching session, we did not show the raw output force data, and the participants were asked to maintain a constant pinch with both their left and right hands simultaneously for 5~s. 
On their left hand, participants maintained the trained 5.0~N. 
On the right hand, participants recreated and matched the perceived pinching force of the left hand. 
As soon as the sensor detected the participants pinching their left hand, we vibrated the left index finger using one of the 4~vibration settings~(\textit{None}, \textit{Flexion}, \textit{Extension}, \textit{Both}) randomly for 5~s. 
After completing a matching session, participants repeated the training session. 
After experiencing the matching session for all 4~vibration settings, participants took a minute break and then repeated the session for a total of 5~times. 
We used the first session as practice.

\begin{figure}[t]
  \centering
  \includegraphics[width=\linewidth]{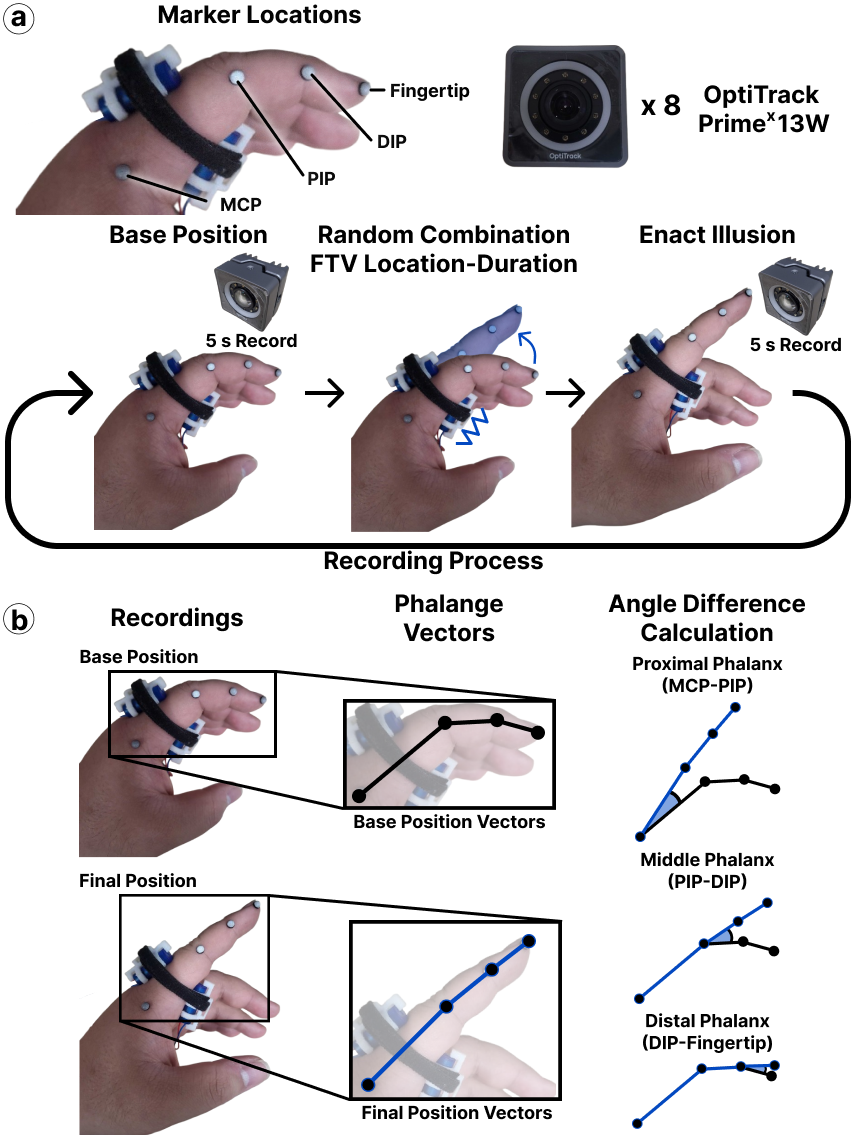}
  \caption{Perception Study~2 setup. (a)~We attached motion tracker markers on the MCP, PIP, and DIP joints and fingertips. Using 8 motion capture cameras, we recorded how participants perceived the FTV-induced movement illusion. (b)~To compute perceived finger movement, we form phalange vectors using joints and fingertip markers. Then, we calculate the rotation angles of the proximal, middle, and distal phalanges.}
  \label{fig:movement setup}
\end{figure}

\subsubsection{Results}
We obtained a total of 112~($28\times4$) pairs of left- and right-hand force data for each vibration setting. 

We first subtracted the left-hand force from the right-hand force to compute the perceived force difference~(Figure~\ref{fig:pinch}b). 
We then divided the 5~s recording into 10 windows of 0.5~s and calculated the average force for each window. 
To see the relative force changes, we subtracted the first average from the rest of the 9 window averages. 
We conducted a one-way analysis of variance~(ANOVA) on each window using the vibration setting as the factor.

Unlike our hypothesis, there was no significant difference between the vibration settings in relative force changes for any of the 0.5~s-windows. 
Since we observed no significant difference, we calculated the effect size using Cohen's f for all windows. We found that all effect sizes were less than medium except for 2 cases: \textit{Flexion} vs \textit{Both} for window 0.5~s$\sim$1.0~s and 4.5~s$\sim$5.0~s. 
Although the effect sizes for these two cases were high (f>0.4), they only appeared at the very beginning and end time windows. 
From this, we conclude that short-duration FTV of less than 5~s does not have a significant effect on perceived voluntarily exerted force. 
As such, FTV-induced illusions would not be appropriate for interactions that utilize voluntarily exerted force such as crushing a virtual object with less actual force. 
For the detailed information about results, including all windows and effect sizes, refer to Appendix~\ref{appendix:perception1}.

\begin{figure}[t]
  \centering
  \includegraphics[width=\linewidth]{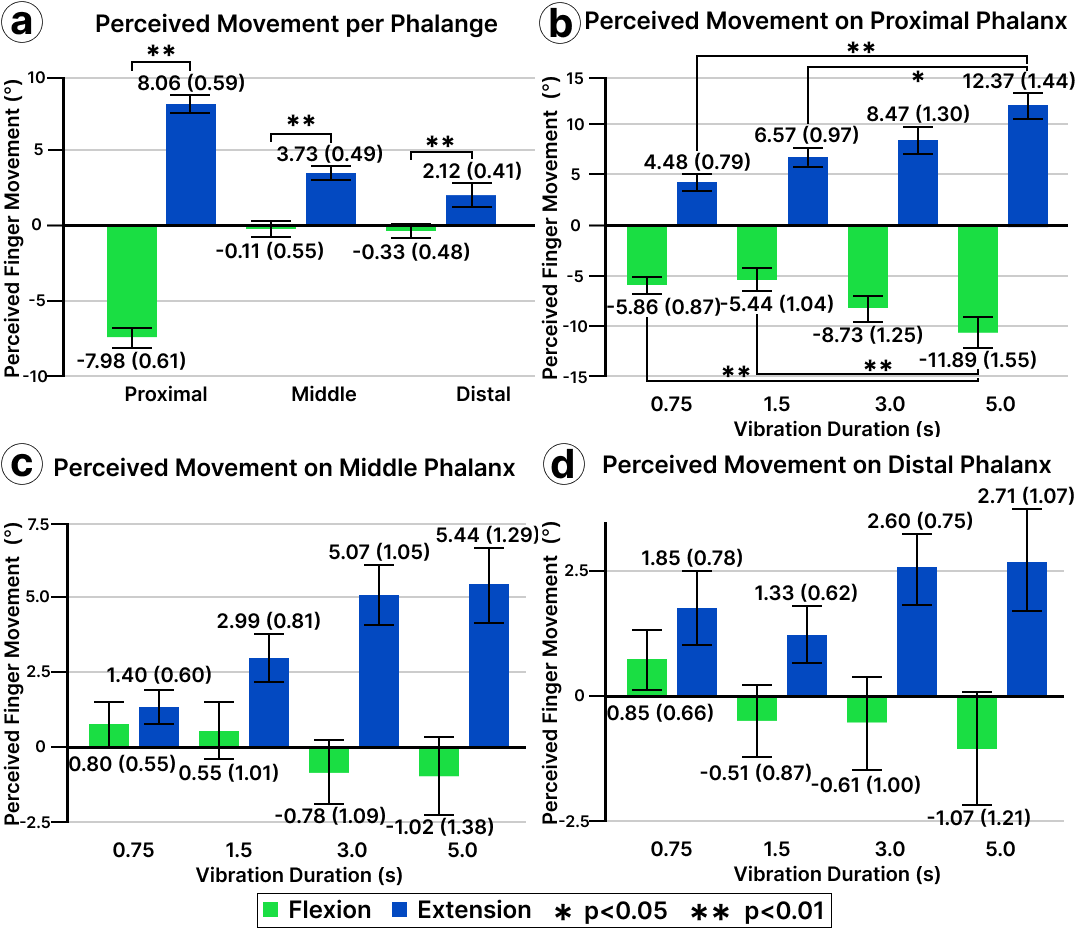}
  \caption{Perception Study 2 results. (a)~We observed clear direction distinction for short-duration FTV. (b)~The perceived proximal phalanx movement had a significant difference between 0.75~s, 1.5~s, and 5.0~s for both flexion and extension. The middle~(c) and distal~(d) phalanges did not show increased perceived movement as vibration duration increased. Numbers above the bar indicate average and standard error, which is also shown by the error bars. * indicates p<0.05 and ** indicates p<0.01.}
  \label{fig:movement}
\end{figure}

\subsection{Perception Study 2: Effect on Perceived Involuntary Movement}~\label{sec:perception2}
\label{sec: perception study 2}
To find the relationship between FTV and perceived movement, we conducted a study where participants self-reported the perceived movement by physically moving their left index finger to the position after the perceived involuntary movement~\cite{sittig1985separate}.

To measure the finger movement, we used 8 motion capture cameras~(Prime$^\text{X}$ 13W, OptiTrack) at 120 frames per second to capture the participants' recreation of the perceived finger movement. 
We attached 4~mm reflective markers on the participants' left index metacarpophalangeal~(MCP) joint, proximal interphalangeal~(PIP) joint, distal interphalangeal~(DIP) joint, and fingertip~(Figure~\ref{fig:movement setup}a).

\subsubsection{Procedure}
After the break from the first perception study, we instructed the participants to relax their left index finger so that the proximal phalanx was around the midpoint of complete extension and flexion. 
Using this position as the base position, we recorded the base position for 5~s and then applied FTV with a random combination of FTV location~(extension, flexion) and FTV duration~(0.75~s, 1.5~s, 3.0~s, 5.0~s). 
After the vibration ended, the participants recreated how they perceived their left index fingers moved from the FTV. 
After they finished moving, we recorded the final position for 5~s. We then instructed the participants to return to the base position and repeated the process for the rest of the location-duration combinations. 
The participants repeated this block of 8 combinations 5~times. Before starting, participants practiced the recording process~(Figure~\ref{fig:movement setup}a) without vibration.

\subsubsection{Results}
We obtained a total of 140~($28\times$5) recordings for each FTV location-duration combination.
From the recordings, we obtained the (X, Y, Z) positions of the MCP, PIP, DIP joints, and fingertip at the base position and final position using the average position of the recorded 5~s~(Figure~\ref{fig:movement setup}b). 
We then obtained the vector values for the proximal phalanx~(MCP-PIP), middle phalanx~(PIP-DIP), and distal phalanx~(DIP-fingertip). 
We compared the phalange vectors at the base position and final position to obtain the rotation angle of each phalanx. However, to account for the finger phalange hierarchy, we calculated the rotation matrix of the proximal phalanx and applied it to the base position's middle phalanx to obtain the rotation angle of only the middle phalanx. 
Likewise, for the distal phalanx, we used the rotation matrix of the proximal and middle phalange. 
For this calculation, we set the extension direction as positive and the flexion direction as negative angles. 

We conducted a two-way ANOVA on each phalange movement data using FTV location and FTV duration as two factors~(Figure~\ref{fig:movement}). 
We used Tukey pairwise comparisons for post-hoc analysis. 

For the proximal phalanx~(Figure~\ref{fig:movement}b), there was a significant difference in perceived movement by FTV location~(F(1, 1112)=368.6, p<0.001) and the interaction effect~(F(3, 1112)=13.56, p<0.001). 
For the flexion direction, we found significant differences between the durations 0.75~s-5.0~s~(p=0.007) and 1.5~s-5.0~s~(p=0.003). 
Likewise for extension direction, 0.75~s-5.0~s~(p<0.001) and 1.5~s-5.0~s~(p=0.012) had significant differences. 
There were significant differences in perceived movement by FTV location for both middle~(F(1, 1112)=27.49, p<0.001) and distal~(F(1, 1112)=15.17, p<0.001) phalanges. 
However, there were no significant differences between perceived movement by FTV duration for flexion or extension.

From these results, we saw that participants felt a clear distinction with perceived movement direction even for FTV durations less than 5.0~s~(Figure~\ref{fig:movement}a). 
Furthermore, for the proximal phalanx, we observed a significant increase in perceived finger movement angle with increased FTV duration.

The proximal phalanx showed an average perceived movement of $-5.86^\circ$ for flexion and $4.48^\circ$ for extension when applied a 0.75~s FTV. 
The $5^\circ$ movement on the proximal phalanx is the same extent of movement when pressing a keyboard key~\cite{peters2002finger}. 
As such we conclude that short-duration FTV evokes large enough illusions to apply to virtual object interactions.

\section{Haptic Rendering for Finger Tendon Vibration in VR}
To summarize the FTV perception study results: 1)~FTV requires a minimum vibration duration of 0.75~s to be effective; 2)~illusion from FTV does not significantly affect perceived voluntarily exerted force; and 3)~illusory involuntary movement from FTV shows a clear distinction depending on stimuli location, which increases with added duration. 
In this section, we utilize our findings to derive 6 interactive VR scenes using FTV that exemplify key VR interactions. 
We also devised different vibration triggers and duration control methods to render FTV using the results.  

We used the head-mounted display~(HMD, Meta Quest 3) and developed exemplary VR scenarios through Unity 3D~(2022.3.36f1). 
To develop rendering methods, we used the OpenXR plugin for Unity 3D. 

\begin{figure*}[t] 
  \centering
  \includegraphics[width=\linewidth]{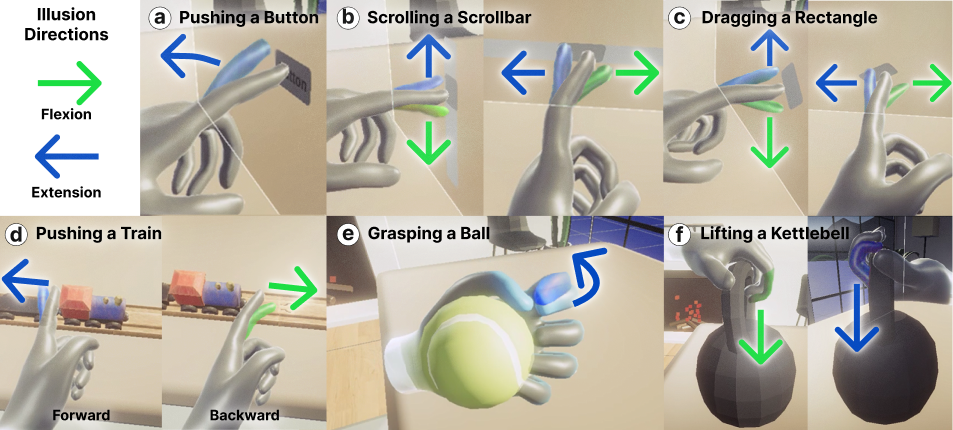}
  \caption{6~exemplary VR scenarios using FTV-induced illusions for immersive virtual object interactions. Green arrows indicate flexion FTV, and blue arrows indicate extension FTV. The colored finger illustrates the illusion the user feels. (a)~The button pushes the finger back. (b)~The resisting force pushes the finger to the opposite direction of the movement. (c)~An opposite force pushes onto the finger. (d)~The user feels the friction as the train is pushed. (e)~The surface normal force pushes the fingers back. (f)~The weight pushes the finger downward.}
  \label{fig:scenarios}
\end{figure*}

\subsection{Exemplary VR Object Interactions} 
Based on the results from the FTV perception study, we define a design space for potential exemplary VR scenarios. 
First, the interaction must be longer than 0.75~s. 
Second, the interaction should not include scenarios that affect user's voluntarily exerted forces. 
Lastly, examples should be interactions that cause involuntary movement or reactionary forces on the finger that would cause the finger to move. 

Based on this, we developed 6~exemplary VR interaction scenarios that cover diverse aspects of VR object interactions from UI components to 3D objects~(Figure~\ref{fig:scenarios}). 
Here, UI components represent point-based, 1D, and 2D interactions, while 3D objects represent experiencing physical properties such as friction, shape, and weight sensations. 
Since hand movement and orientation determine how the finger will interact with the virtual objects, we fix the hand movement and posture where the left palm either faces downwards or in the medial direction, unless specified otherwise. 

\textbf{\textit{Button~(Point-based)}.} 
For a point interaction UI component, we choose a virtual button~(Figure~\ref{fig:scenarios}a). 
Using a real button as a metaphor, there is a constant reaction force from the button, which pushes the index finger back with the extension illusion. 
To simulate the finger movement, we apply FTV to generate a sensation of finger extension. 
While in this paper, we only focus on a singular button using the left index finger, this can be expanded to keyboards by applying FTV to several fingers. 

\textbf{\textit{Scrollbar~(1D)}.}
For a 1D UI component interaction, we devise a scrollbar as an example(Figure~\ref{fig:scenarios}b). 
We provide a resistance force against the scrollbar handle movement that pushes against the finger to deliver compliance. 
For the vertical scrollbar, we used FTV to elicit extension for a downward handle movement and flexion for an upward movement. 
For the horizontal scrollbar, we used FTV to elicit extension for the handle moving to the right and flexion FTV for the handle moving to the left.

\textbf{\textit{Rectangle~(2D)}.}
For a 2D UI component manipulation scenario, the user selects a 2D rectangle object and drags it across the screen~(Figure~\ref{fig:scenarios}c).  
Like \textit{Scrollbar}, we apply FTV as a resisting force against the movement direction pushing on the finger. 
But unlike \textit{Scrollbar}, where the direction of FTV perfectly aligns with the movement, \textit{Rectangle} moves in a 2D space where the direction of the FTV illusion only partially aligns with the movement. 
We designed two drag trajectories for \textit{Rectangle} with a diagonal path: a rectangle that moves from the top-left to the bottom-right and a rectangle from the bottom-left to the top-right. 
For the former rectangle, we apply extension FTV when the rectangle moves downward and flexion when it moves up. 
For the latter rectangle, when the hand moves the rectangle towards the top-right, we apply extension FTV, and when the hand moves the rectangle to the bottom-left, we apply flexion FTV.

\begin{figure}[t] 
  \centering 
  \includegraphics[width=\linewidth]{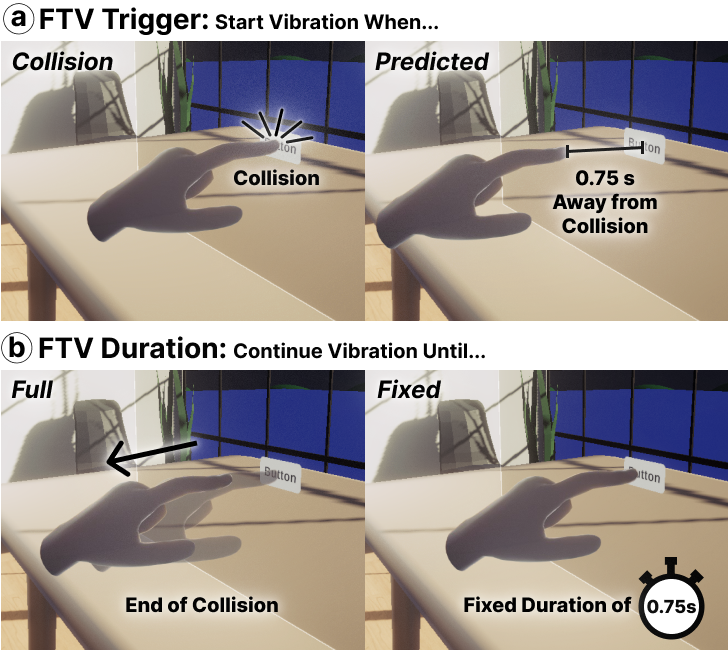}
  \caption{FTV rendering methods. (a)~We implemented two rendering methods to start FTV, where \textit{Collision} starts vibrating when a collision is detected and \textit{Predicted} starts vibrating at a distance that would take 0.75~s to collide with the target object. (b)~We also implemented two duration control methods where \textit{Full} continues the vibration until the end of the collision and \textit{Fixed} only vibrates for a fixed duration of 0.75~s.}
  \label{fig:rendering methods} 
\end{figure}

\begin{figure*}[t]
    \centering
    \includegraphics[width=\linewidth]{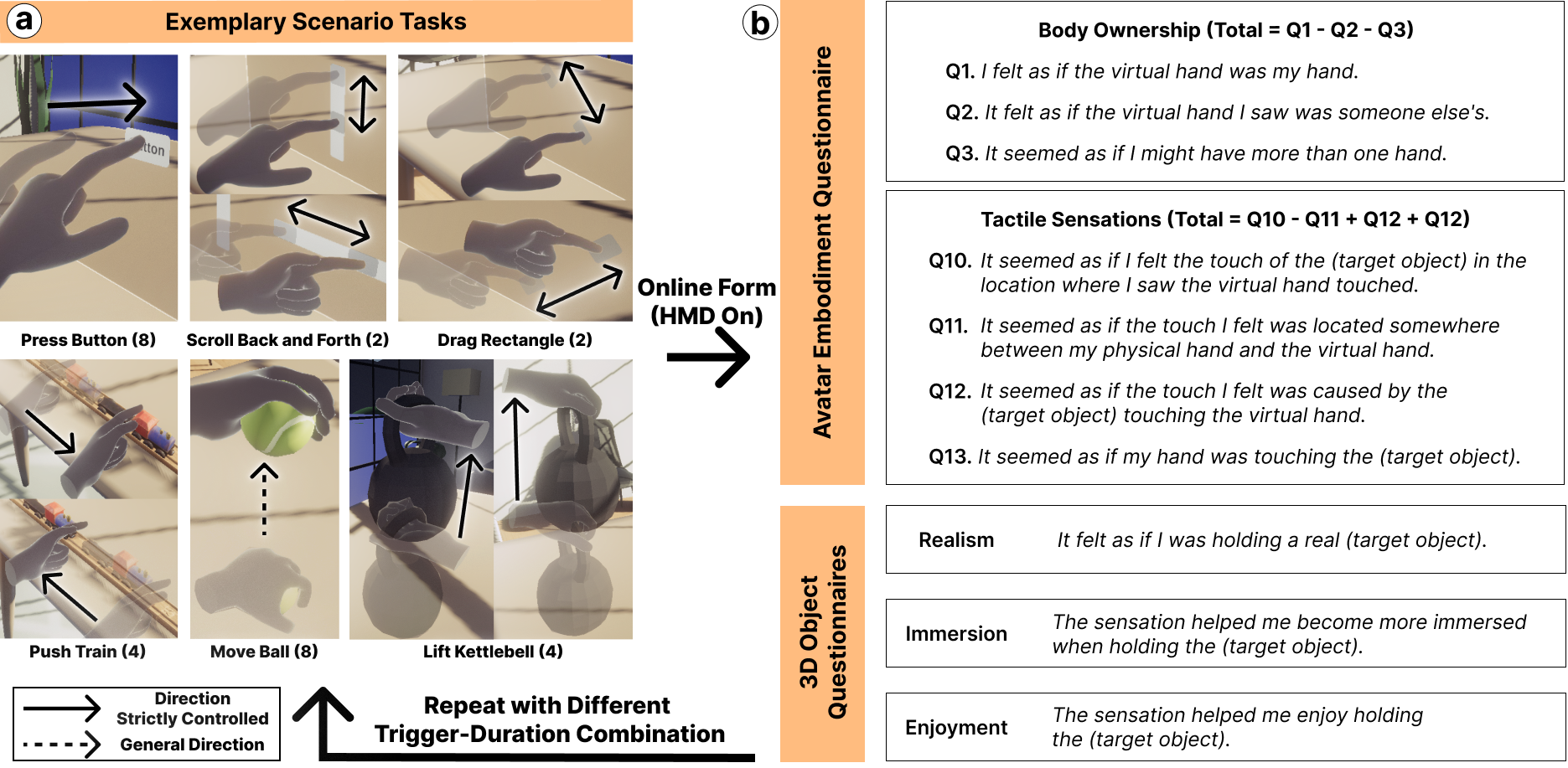}
    \caption{Rendering evaluation procedure. After completing the task for each exemplary scenario, participants answered the questionnaire through an online form in VR. After completing the form, participants repeated the process for a different FTV rendering setting. (a)~Task for each exemplary scenario. The number in parentheses indicates task repetition. (b)~Questionnaires used in the evaluation. For each scenario, the \textit{(target object)} was changed to fit the scenario. We include the arithmetic formula and question numbers listed in the avatar embodiment questionnaire.}
    \label{fig: eval procedure}
\end{figure*}

\textbf{\textit{Train~(Friction)}.}
The next three exemplary interactions demonstrate the experience of diverse physical properties with 3D objects. 
We choose a train~(Figure~\ref{fig:scenarios}d) to imitate friction using FTV. 
Unlike the virtual UI components that generate a constant force throughout the interaction, there exists a larger initial force as the train moves from the static friction. 
In this example, we designed a scenario where the user would have their left palm facing the medial direction and push the train using the index finger.
When pushing the train forward, we use extension FTV to recreate the reactionary force from friction. 
When pushing the train backward using the dorsal side of the index finger, we use flexion FTV. 

\textbf{\textit{Ball~(Shape)}.}
When grabbing an object, the normal force from the object's surface pushes against the hand and finger, providing spatial information on the shape. 
Since we only apply FTV to the index finger, we focus on the normal force sensation of an object pressing against the finger, instead of detailed shape recognition. 
For this example, we chose a tennis ball~(Figure~\ref{fig:scenarios}e), which can be easily grabbed with a single hand. 
Unlike other exemplary scenarios, we do not have a fixed hand movement direction for this scenario, as holding the ball is sufficient to feel the normal force. 
Since the normal force only pushes the finger outwards, we only apply extension FTV for \textit{Ball}.

\textbf{\textit{Kettlebell~(Weight)}.}
To simulate weight using FTV, we chose a kettlebell~(Figure~\ref{fig:scenarios}f). 
Holding a kettlebell creates a constant downward force because of the weight.
Unlike \textit{Ball}, where the grip force is constantly outwards the object, we use FTV illusions for the constant downward force for \textit{Kettlebell}.
The direction of the force on the finger depends on whether the palm is facing up or down. 
If the palm is facing down, the weight of the kettlebell handle will pull the fingertip downwards. 
Here, the proximal phalanx moves in the flexion direction, but the middle phalanx opens up in the extension direction. 
However, since we observed how FTV affects the proximal phalanx more, we apply flexion FTV to simulate the weight. 
On the other hand, if the palm is facing up, the kettlebell handle presses the finger back, causing an extension in the direction of movement. 
We simulate this movement sensation using flexion and extension FTV for their respective cases.  

\subsection{FTV Rendering Methods}~\label{sec:rendering_methods} 
Here, we detail when and for how long FTV will be activated in a VR object interaction scenario. 
For each vibration trigger~(Figure~\ref{fig:rendering methods}a) and vibration duration~(Figure~\ref{fig:rendering methods}b), we devised additional rendering methods that use the perception study results. 

\subsubsection{FTV Trigger} 
\textbf{\textit{Collision}.} 
For this method, we follow conventional haptic rendering methods and start applying FTV when the system recognizes the left index finger colliding with the object. 
For UI components, we use the built-in Poke interaction from OpenXR. 
For \textit{Train}, we use a collider on the left index fingertip. For \textit{Ball} and \textit{Kettlebell}, we use the built-in grasp recognition from OpenXR. 

\textbf{\textit{Predicted}.}
From the preliminary perception study, we found that FTV requires a minimum of 0.75~s of vibration to perceive the illusion. 
In \textit{Predicted}, we take this minimum duration into account and start vibrating 0.75~s before the user touches the object. 

To calculate when to activate the vibrations, we conducted a pilot study with 10 participants~(5~male, 5~female; mean age$=25.0$, SD$=2.73$). 
For each exemplary scene, we asked participants to steadily and naturally move their left hand toward each target object. 
We logged when the left index fingertip was 10~cm, 8~cm, 6~cm, 4~cm, and 2~cm away from the target and when the fingertip collided with the object. 
Using these data points from the 10 participants, we used linear regression to calculate the distance of the finger 0.75~s before the collision. 
Based on this, we used 4.88~cm as the vibration-triggering distance for \textit{Predicted} trigger.

\subsubsection{FTV Duration}

\textbf{\textit{Full}.}
For rendering vibration for virtual object interactions, the vibration continues as long as the user is in contact with the virtual object. 
We refer to this as \textit{Full} and stop the vibration when the user loses contact with the object.

\textbf{\textit{Fixed}.}
From the results of the second perception study~(Section~\ref{sec: perception study 2}), we concluded that FTV durations between 0.75~s and 1.5~s do not have significant differences in perceived movement. 
Using this, for object interactions between 0.75~s and 1.5~s, vibrating the tendon for 0.75~s should result in similar proprioceptive sensations. 
As such, we propose a \textit{Fixed} duration method for rendering FTV where a fixed duration of 0.75~s is applied for short interactions of less than 1.5~s. 
We hypothesize that for these interactions, the illusory movement from 0.75~s FTV should be enough for all interactions of that length. 
As long as \textit{Fixed} provides a similar experience as \textit{Full}, \textit{Fixed} has an advantage in simplicity and power consumption. 

\section{Rendering Method Evaluation}~\label{sec:rendeirng_eval}
In this section, we explore how to most effectively render FTV for VR object interactions.
We apply FTV rendering methods based on the results from our perception study to our exemplary VR scenarios. 
These methods~(\S\ref{sec:rendering_methods}) include two different methods for each FTV trigger and duration.
Since FTV requires a minimum duration of 0.75~(\S\ref{sec:perception2}), we developed a \textit{Predicted} FTV trigger and compared it with the conventional \textit{Collision} trigger.
Since there is no significant difference in perceived movement for FTV durations between 0.75~s and 1.5~s~(\S\ref{sec:perception2}), we developed a \textit{Fixed} duration for FTV, hypothesizing that it would have no difference compared to the conventional \textit{Full} duration for interactions between 0.75~s and 1.5~s.
With a total of 4~possible FTV rendering method, we apply these rendering methods to all VR scenarios to compare which rendering method most enhances the VR experience.

\subsection{Evaluation Design}
We recruited 16~participants~(8~male, 8~female; mean age$=24.9$, SD$=1.5$) who were all right handed. 
Participants experienced each exemplary VR scenario with each combination of vibration trigger and vibration duration~(Figure~\ref{fig: eval procedure}). 

\textbf{Questionnaires.}
We utilized parts of the avatar embodiment questionnaire relevant to our evaluation~\cite{gonzalez2018avatar}~(Figure~\ref{fig: eval procedure}b). 
Out of the six categories, we only used questions on body ownership and tactile sensations. 
Since our VR scene does not contain any mirror, we used only questions Q$1\sim3$ in the original questionnaire for body ownership. 
For tactile sensations, we used all 4 questions~(Q$10\sim13$) in the original questionnaire. 
For the total body ownership and tactile sensation score, we used the formula noted in the embodiment questionnaire~\cite{gonzalez2018avatar}. 
For 3D objects~(\textit{Train}, \textit{Ball}, \textit{Kettlebell}), we added 3~questionnaires related to realism, immersion, and enjoyment.
We used a 7-point Likert scale for all questionnaires. Each wording of the questionnaire was modified slightly to match the exemplary scenario. 
Figure~\ref{fig: eval procedure}b shows the general questionnaire format. 
For the exact wording of the questionnaires for each exemplary scenario, refer to Appendix~\ref{appendix:questionnaires}. 

\textbf{Procedure.}
Participants experienced the exemplary scenarios in the order introduced in this paper. Each exemplary scenario had a different task~(Figure~\ref{fig: eval procedure}a). 
We set the frequency of each task so that participants would interact with the target object 8~times regardless of FTV direction or hand orientation. For \textit{Button}, participants pressed the button 8~times. 
For \textit{Scrolling}, participants moved the handle in the scrollbar back and forth twice for each scrollbar~(2~scrollbars~$\times$~2~directions~$\times$~2~tries). Likewise, for \textit{Dragging}, participants moved each rectangle diagonally back and forth twice~(2~rectangles~$\times$~2~directions~$\times$~2~tries). For \textit{Train}, participants moved the toy train up and down the track 4~times~(2~direction~$\times$~4~tries). For \textit{Ball}, participants grasped the ball, moved freely, and released the ball 8~times. For \textit{Kettlebell}, participants lifted each kettlebell and released them 4~times(2~kettlebells~$\times$~4~tries). When interacting with each target object, we instructed the participants to keep the interaction between 0.75~s to 1.5~s to compare the effectiveness between \textit{Full} and \textit{Fixed}. Participants could check how long they interacted with the object by checking a stopwatch displayed in front of them. 

To prevent VR inexperience from biasing the result, participants experienced the scenario without any vibrations before starting an exemplary scenario until they were proficient with the interaction, which took no longer than 5~minutes. 
Within each exemplary scenario, participants experienced one of the four FTV trigger-FTV duration combinations in random order. 
After completing a combination, participants answered the questionnaire through an online form in VR. 
After finishing all 4~combinations for an exemplary scenario, the participant removed the HMD, ranked the combinations by their preference, and took a 2~minute break before repeating the process for the next exemplary scenario. 
After completing all exemplary scenarios, we conducted a free-form interview for comments on how they experienced each exemplary scenario. 
The whole evaluation took no longer than 1.5~hours.

\subsection{Results}
From this study, we aimed to observe which rendering method is most effective for the exemplary VR scenarios we prepared. 
Since all the questionnaire results are ordinal data, we use a two-way aligned rank transform analysis of variance~(ART ANOVA) to find significant differences between rendering methods.

\begin{figure*}[t]
  \centering
  \includegraphics[width=\linewidth]{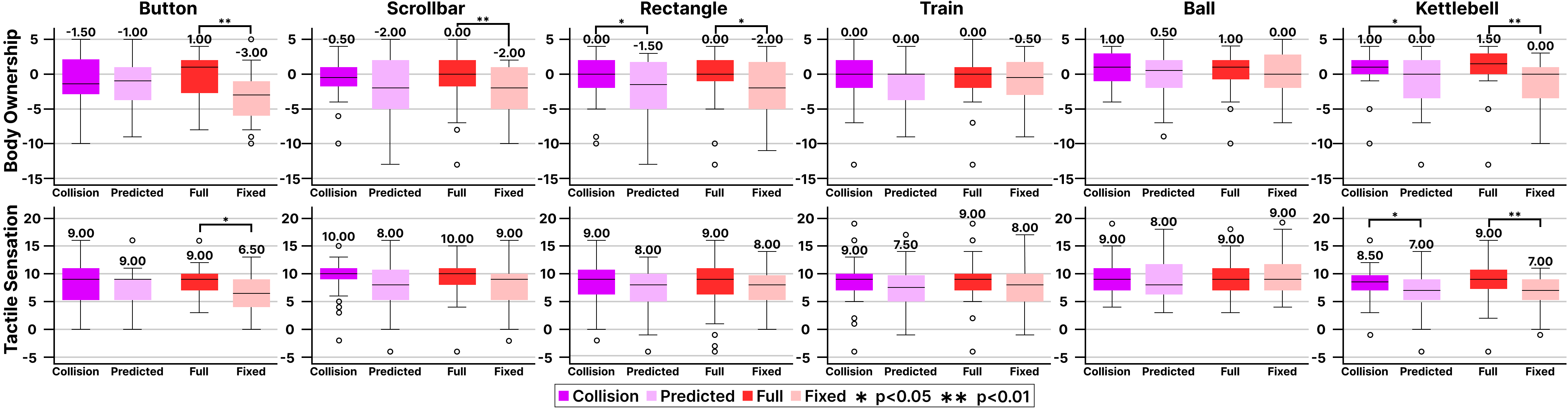}
  \caption{Boxplot of body ownership and tactile sensation answers from the avatar embodiment questionnaires. Circles denote outliers, and the median is shown on top of the boxplot. * indicates p<0.05 and ** indicates p<0.01.}
  \label{fig:embodiment results}
\end{figure*}

\textbf{Avatar Embodiment Questionnaire Results.}
We used the score calculation formula from the avatar embodiment questionnaire~(Figure~\ref{fig: eval procedure}b)~\cite{gonzalez2018avatar}. 
Since we used a 7-point Likert scale, the body ownership scores ranged between $-13\sim5$ and tactile sensation ranged between $-4\sim20$. 
Because no significant interaction effect was observed, we report our findings by comparing only trigger and duration methods~(Figure~\ref{fig:embodiment results}).

Out of all scenarios, \textit{Train} and \textit{Ball} showed no significant differences in rendering method. 
\textit{Button} showed significant increase in body ownership~(F(1, 60)=7.87, p=0.007) and tactile sensation~(F(1, 60)=6.24, p=0.015) for \textit{Full}. 
For \textit{Scrollbar}, \textit{Full} had a significant increase in body ownership~(F(1, 60)=7.29, p=0.009). 
For \textit{Rectangle}, both trigger method~(F(1, 60)=4.09, p=0.048) and duration method~(F(1, 60)=4.14, p=0.046) had significant differences for body ownership. 
\textit{Kettlebell} had significant differences for both body ownership and tactile sensation. 
On body ownership, \textit{Collision} had a significant increase~(F(1, 60)=4.15, p=0.046) along with \textit{Full}~(F(1, 60)=7.60, p=0.008). 
For tactile sensation, \textit{Collision}~(F(1, 60)=4.79, p=0.032) and \textit{Full}~(F(1, 60)=7.83, p=0.007) similarly had a significant increase.

We hypothesize that the difference between scenarios that had significant differences in body ownership for duration methods is because of differing levels of constant touch. 
For \textit{Button}, \textit{Scrollbar}, \textit{Rectangle}, and \textit{Kettlebell}, there is a constant level of force involved during the interaction. 
However, for \textit{Train}, the initial push is when the force is highest and then drops. 
For \textit{Ball}, as a person moves their hand, the grip forces change depending on the hand orientation~\cite{slota2011grip}. 
Considering how sensory expectation affects body ownership~\cite{ferri2013body}, the additional vibrational tactile feedback in \textit{Full} may have supported participants' expectations during scenarios with a constant force level. 
On the other hand, for scenarios with fluctuating forces such as \textit{Train} and \textit{Ball}, since participants do not expect a constant force, we hypothesize that the lack of additional tactile feedback did not harm the experience during \textit{Fixed}.

Meanwhile, only \textit{Button} and \textit{Kettlebell} maintained a significant difference between duration methods for tactile sensations out of scenarios that showed a significant difference in body ownership. 
Although we also delivered a constant resisting force against the hand movement for \textit{Scrollbar} and \textit{Rectangle}, these scenarios may not have shown much difference for tactile sensation because the direction of resistance against the finger. 
In these two scenarios, the finger extension and flexion movement, which FTV illusion provides, are not the direct result of the movement of the scrollbar handle and the rectangle. 
For the other 4~scenarios the forces acting on the finger are on the same axis as the target movement, directly causing finger extension and flexion movement.
On the other hand, for \textit{Scrollbar} and \textit{Rectangle}, the extension and flexion finger movement is an indirect result that occurs from the resistive force acting upon the fingertip surface during lateral hand movement.

\textbf{Feedback from Participants.} 
As FTV's movement illusion is an unfamiliar feeling, many participants noted how the sensation felt weird and new. 
``It was interesting. I never felt something like this before''~(P11). 
However, some participants noted how it felt unnatural for \textit{Fixed}. 
``It was an interesting feeling when the feeling was maintained while touching the object"~(P6). 
``It was weird when the vibration stopped early for \textit{Button}''~(P8). 
Another participant noted how it might have been better if they could feel the vibration on all fingers. 
``It felt realistic until \textit{Train} because I only used my index finger. For \textit{Ball} and \textit{Kettlebell}, they didn't feel as realistic"~(P12). 
An interesting observation was how the visual information during VR may have affected participants' FTV perception. 
``The finger movement felt weaker in VR than in calibration''~(P14).

\begin{figure*}[t]
  \centering
  \includegraphics[width=\linewidth]{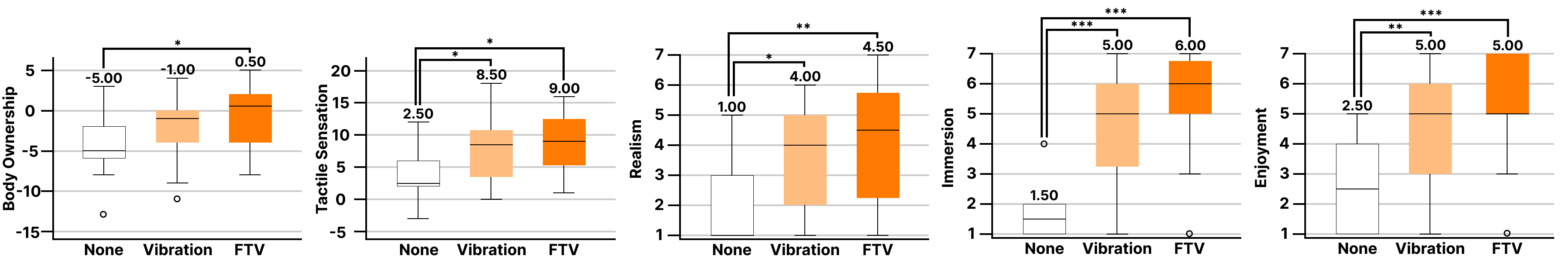}
  \caption{Boxplot questionnaire results. Circles denote outliers, and the median is shown on top of the boxplot. * indicates p<0.05, ** indicates p<0.01, and *** indicates p<0.001.}
  \label{fig:new eval results}
\end{figure*}

\section{VR Experience Evaluation}
In this section, we evaluate the effect of FTV on VR experience. 
Using the results from our rendering method evaluation~(\S\ref{sec:rendeirng_eval}), we fix the rendering method for FTV to \textit{collision} and \textit{Full}.
We compared this against two baseline methods, no vibration and simple vibration, using our exemplary VR scenarios. 
Throughout this evaluation, we confirm that FTV facilitates an enhanced VR experience with superior body ownership.

\subsection{Evaluation Design}
We recruited 16~participants~(6~female; mean age$=25.4$, SD$=3.6$) who were all right-handed. 
We used the same exemplary VR scenarios used for the rendering method evaluation. We also used the same questionnaires as the evaluation in \S\ref{sec:rendeirng_eval}~(Figure~\ref{fig: eval procedure}b). 
Participants experienced the VR scenarios using 3~different haptic settings: \textit{None}, \textit{FTV}, and \textit{Vibration}. For \textit{None}, there were no vibrations. For \textit{FTV}, we delivered FTV following the \textit{Collision} and \textit{Full} rendering methods. 
For \textit{Vibration}, we vibrated both ERM motors simultaneously on the finger when participants touched virtual objects. 
By vibrating both sides of the finger, we aimed to cancel out the FTV illusion that would be caused by the vibrations.
This way, we prepared a vibrotactile feedback for \textit{Vibration} that has the same vibration specification of 80~Hz and 1~mm vibration displacement amplitude and location as \textit{FTV}, but does not elicit the finger movement illusion.

\textbf{Procedure.} Participants freely experienced the VR scenarios in any order for at least 3~minutes by clicking on a menu on the side, which spawned the respective VR objects for each scenario. 
After experiencing the VR scenarios for one setting, the participants answered the questionnaire.
After experiencing all exemplary VR scenarios for all settings, we interviewed the participants, asking which setting was their favorite and why, which VR scenario was their favorite and why, and lastly, if there were any aspects of the experience they found comfortable or uncomfortable. 
The whole evaluation took no longer than 45~minutes.

\subsection{Results}
While the participants were free to choose the order they experience the VR scenarios, most chose to follow the order that was shown on the menu~(\textit{Button}, \textit{Scrollbar}, \textit{Rectangle}, \textit{Train}, \textit{Ball}, \textit{Kettlebell}).
We used a Kruskal-Wallis test and a post-hoc Dunn's test to compare the results for the questionnaires~(Figure~\ref{fig:new eval results}). 

\textbf{Questionnaire Results.}
Similar to our rendering method evaluation, we used the score calculation formula from the avatar embodiment questionnaire~(Figure~\ref{fig: eval procedure}b)~\cite{gonzalez2018avatar}, resulting in a range of $-13\sim5$ for body ownership and $-4\sim20$ for tactile sensation. There was a significance difference between vibration settings for body ownership~(H(2)=6.566 p=0.038), tactile sensation~(H(2)=6.778 p=0.034), realism~(H(2)=9.944 p=0.007), immersion~(H(2)=22.643 p<0.001), and enjoyment~(H(2)=14.446 P<0.001).

For body ownership, only \textit{FTV} had a significant increase from \textit{None}~(p=0.011). 
For tactile sensation, both \textit{FTV}~(p=0.014) and \textit{Vibration}~(p=0.049) had significant increase. 
Likewise, for realism, both \textit{FTV}~(p=0.003) and \textit{Vibration}~(p=0.021) had significant increase.
Immersion showed a more profound increase where both \textit{FTV}~(p<0.001) and \textit{Vibration}~(p<0.001) had significant increase.
Similarly for enjoyment, both \textit{FTV}~(p<0.001) and \textit{Vibration}~(p=0.006) had significant increase.
From the 5~criteria, only body ownership had \textit{FTV} show a significant increase. 
Here, we can conclude that vibrotactile feedback is enough to improve the other four criteria, whereas FTV's finger movement illusion was essential for improving body ownership. The underlying reason for this result becomes more evident with the interview results.

\textbf{Interview Results.}
For the vibration setting preference, 2 answered \textit{None}, 4 answered \textit{Vibration}, and 10 answered \textit{FTV}.
Participants who favored \textit{None} wanted to imagine the feeling by themselves and refused haptic feedback. 
Participants who preferred \textit{None} reported that they felt most familiar with the vibrotactile feedback from \textit{Vibration}.
Lastly, participants who preferred \textit{FTV} commonly referred to the feeling of resistance with clear direction on their finger with the following reasons.
``I did not feel a sense of grasping or force with \textit{None} and \textit{Vibration}, but I felt it with \textit{FTV}''~(P3).
``It felt like an actual physical resistance against the direction of my movement. It felt realistic.''~(P15).

This feeling of resistance with direction was also evident in the participant's preference for VR scenarios.
\textit{Train} was most preferred with 5 votes.
Participants mainly commented on how the force they felt against their finger when pushing the train felt realistic.
``I liked the train because I felt what I would expect when I push a toy train with my finger''~(P14).
Surprisingly, \textit{Ball} and \textit{Kettlebell} had the most varied response, with 4 participants choosing it as their favorite because they felt like they were grasping an object.
``The kettlebell felt more difficult to lift''~(P4).
However, there were also participants who reported that FTV was not enough to feel a sense of grasping.
``I didn't feel the ball that well. It didn't feel like I was holding something''~(P6).
The individual differences in perceiving the illusion may have played a role in the grasping task that involves all fingers. 
Those who felt the illusion strongly felt that the illusion on the index finger was enough for the sense of grasping. On the other hand, participants who barely feel the illusion may require FTV over multiple fingers.

While the participants did not report discomfort regarding the motor placement on the finger, they mentioned occasional hand tracking loss due to the occlusion caused by the motor. 
An interesting observation was how the visual information amplified participants' experience of the FTV illusion.
``I felt the push on my finger much more when there was visual information compared to during the calibration''~(P5).
However, some participants also mentioned that the illusion would have felt more realistic if it had become stronger proportionally to the force they applied with their finger.

\section{Discussion}
\subsection{FTV for Immersive Virtual Object Interactions}
Compiling the results of our perception studies, rendering method evaluation, and VR experience evaluation, we outline a design guideline to utilize FTV-induced illusions for immersive virtual object interaction. 
To effectively use FTV, the object interaction should last longer than 0.75~s for users to perceive the illusion. 
For interactions of that length or longer, the illusion onset delay is insignificant. 
However, the tactile cue from longer vibrations, even if the additional duration has no effect on the perceived illusion, is crucial for scenarios with constant force. 
Lastly, VR applications should focus on resistive push against the finger to most effectively employ FTV, and individual differences should be considered when deciding where to apply FTV.

\textbf{Inducing the Illusion.}
Since the minimum duration to evoke the kinesthetic illusion is 0.75~s, FTV should be applied when the simulated forces continue for longer than 0.75~s. 
Although users start to feel the illusion after 0.75~s, \textit{Collision} performed as well and even sometimes better than \textit{Predicted}. 
From this, the 0.75~s delay of proprioceptive cue onset may be negligible as long as users continue the interaction for longer than 0.75~s. 

\textbf{Vibration Duration.}
FTV durations with differences greater than 3.5~s produce different perceived movement illusions. 
Although illusions induced by FTV durations 0.75~s and 1.5~s have no significant psychophysical difference, designers cannot ignore the additional tactile cue given by the vibration itself, which is also supported by participant's preference and greater score for \textit{Collision} compared to \textit{Predicted}.
Even though the additional vibration duration does not affect perceived movement, we observed in the rendering evaluation that vibrational cues reinforce the user's expectation of a constant touch for object interactions with a constant force level. 
We predict that a strategy like cutting the vibration early would be effective in scenarios where the rendered forces are not constant.

\begin{figure}[t]
  \centering
  \includegraphics[width=.8\linewidth]{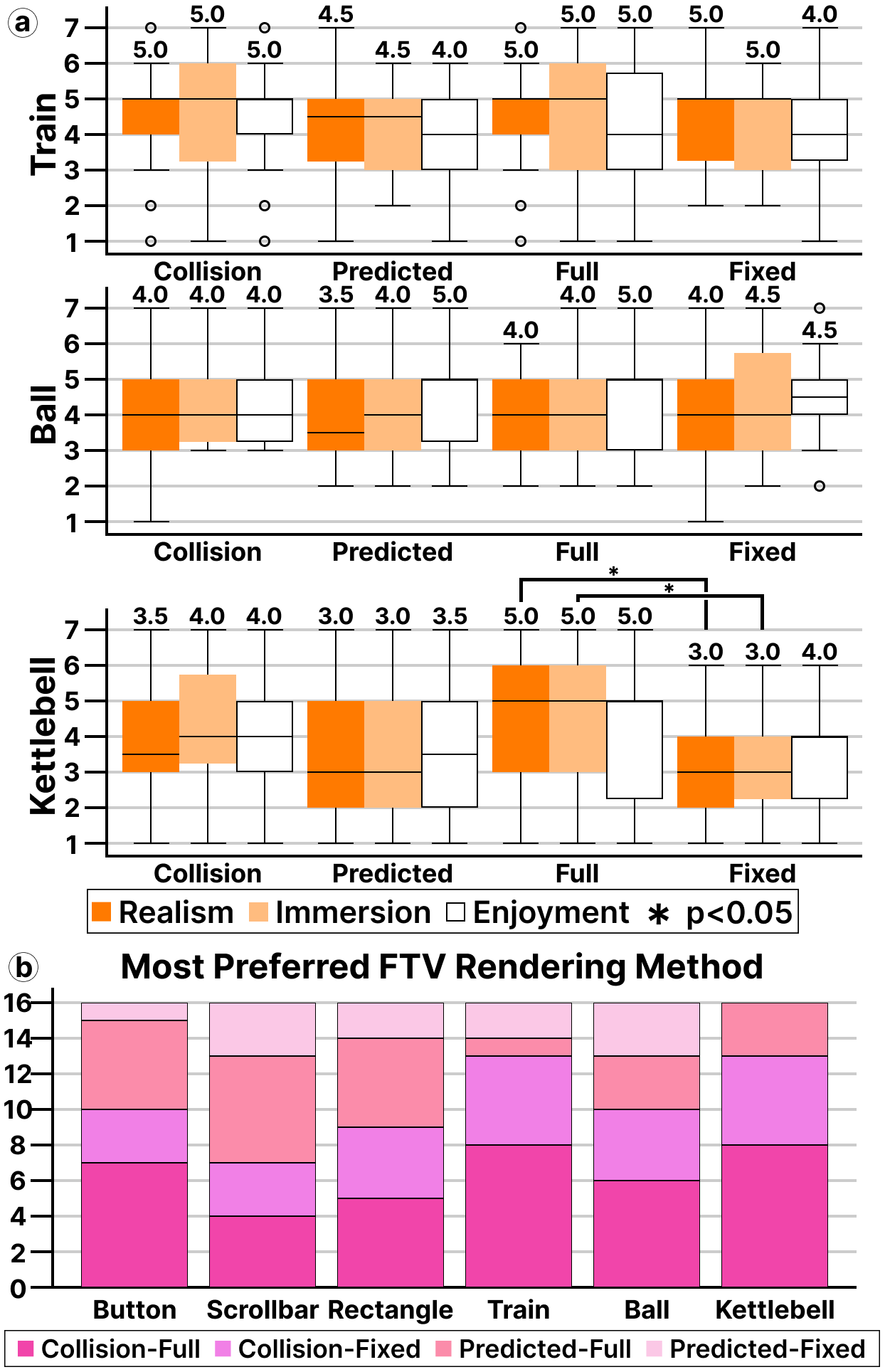}
  \caption{Boxplots on realism, immersion, and enjoyment for (a)~\textit{Train}, (b)~\textit{Ball}, and (c)~\textit{Kettlebell} during the rendering method evaluation. Circles denote outliers, and the median is shown on top of the boxplot. * indicates p<0.05. (d)~Most preferred FTV rendering method for each exemplary scenario.}
  \label{fig:discussion}
\end{figure}

\textbf{Familiarity and Other Metrics.} 
For the additional questionnaires asked for 3D objects, only \textit{Kettlebell} showed significant difference between \textit{Full} and \textit{Fixed} realism~(F(1, 60)=11.7, p=0.001) and immersion~(F(1, 60)=10.7, p=0.002)~(Figure~\ref{fig:discussion}a). 
The different significant results per scenario support the same observations we saw in the avatar embodiment questionnaire. 
However, \textit{Kettlebell}, which showed significant differences in duration methods for body ownership, tactile sensation, realism, and immersion, did not have a significant difference in enjoyment.
Here, we assume that even without tactile cues that supported participants' expectations, FTV's kinesthetic illusion provided enough enjoyment to lessen the score drop compared to the score drop for realism and immersion from \textit{Full} to \textit{Fixed}.
Since the scope of the evaluation was to determine the best FTV rendering method, we did not compare the scenarios.
However, there is room for exploration in how FTV is applied throughout different scenarios, considering the different performance results~(Figure~\ref{fig:embodiment results}).

\textit{Collision}-\textit{Full} consistently ranked as the most preferred FTV rendering method~(Figure~\ref{fig:discussion}d). 
We expected \textit{Full} to perform better than \textit{Fixed} as seen from the questionnaire results. 
However, preferences for rendering methods with \textit{Predicted} were surprisingly lower. 
While the questionnaire focuses on body ownership and tactile sensations, preference also includes different qualitative categories like comfort and familiarity. 
Here, we assume that \textit{Predicted} vibration timing before touch was too unfamiliar for participants. 

\textbf{FTV vs. Simple Vibration.} 
Through the VR experience evaluation, we observed that FTV's greatest strength is its realistic resistance sensation on the finger that simple vibration alone cannot achieve.
However, there are still other factors to consider when comparing FTV and traditional vibrotactile feedback. 
To compare FTV with simple vibration in our work, we kept the frequency and location the same, but for a traditional vibrotactile feedback, the actuator would be attached to where VR users would contact the VR object. 
While we have not explored the effect of vibration on different locations for tendon vibration, further research is needed to observe the perceptual difference between FTV and vibrotactile feedback.
This would lead to more engaging tactile feedback methods involving smoothly transitioning between FTV's kinesthetic illusion and vibrotactile feedback or simultaneously using both.

\textbf{FTV for VR object interaction.}
In the VR experience evaluation, both \textit{Vibration} and \textit{FTV} showed an increase in average body ownership scores compared to \textit{None}, which was expected by previous works that provided haptic feedback in virtual environments~\cite{padilla2010hand}.
However, only \textit{FTV} showed a significant difference with \textit{None}.
This difference can be attributed to its strength in providing a feeling of resistance with direction, as noted by the participants in the interview.
Based on this result, aside from the 6~exemplary VR scenarios showcased in this work, we expect that FTV will best simulate kinesthetic feedback for object interactions that involve a resistive push against the finger.
Also, while FTV can also be used to provide realistic grasping sensations, for our current setup, which only affects the index finger, the sensation may vary largely depending on individual sensitivity towards FTV illusions. 

\textbf{Multi-finger FTV.}
Our current work only included the index finger eliciting motion illusion in the flexion and extension direction. Due to the fixed hardware setup, our VR interaction scenarios only involve one finger. However, many virtual object interactions require the involvement of multiple fingers. On top of that, finger movements are not limited to flexion and extension motions.
To form more realistic feedback, whole-hand FTV with abduction illusion should be implemented. However, we need to consider different finger sensitivities~\cite{sutherling1992cortical} that may result in differing FTV-induced illusion effectiveness and vibration motor locations to avoid hindering hand and finger movement.

\textbf{Additional Applications.} 
Lastly, we highlight prospective applications for FTV in VR. 
Applications, aside from simulating realistic haptic feedback, that require a sensation of resistance and direction would best benefit from FTV.
For example, the finger movement illusion can provide intuitive directional cues through movement illusions toward a target for guidance or navigation. 
FTV illusion can also provide contextual haptic feedback by adding resistance to the finger when moving the ray or pointer.

\subsection{Limitations and Future Works}
To achieve the required vibration parameters, we used a cylinder ERM motor. 
Although the motor did not hinder most participants' movement, we had to omit VR application cases that required a full finger flexion because, depending on finger length, the motor on the palmar side would block the finger from performing a full finger flexion. 
Furthermore, while the ERM delay is still within bounds for most of the population to not notice visual-haptic asynchrony~\cite{di2019perceptual}, the average of 50.5~ms delay is not negligible, especially with shorter durations. 
This is especially crucial as shown in our minimum duration threshold perception study, where many of the participants perceived FTV illusion at the lowest threshold of 0.5~s.
With an actuator with more accurate duration control, there is still room for exploration for even shorter minimum duration with smaller step sizes for FTV illusion elicitation.
Lastly, because of the location of the motor, using the current setup for FTV occasionally causes visual occlusion for other fingers when using visual hand tracking.
This showed the need for further exploration of different types of vibration motors, which are small but strong enough to induce illusions.

One limitation in our perception studies is that the results rely on self-reported measures. 
We modified methods conducted in previous research using tendon vibration~\cite{tidoni2015illusory,reschechtko2018force,sittig1985separate}. 
However, there is still a possibility of noise introduced into the data because of the nature of self-reporting.
While it is possible to use electromyography~(EMG) to measure the effects of tendon vibration~\cite{calvin2000relations}, for FTV, the surface area of the finger is limited for both actuator and EMG electrodes. 

In this paper, we focused on the feasibility of applying short-duration FTV for kinesthetic haptic feedback. 
Hence, throughout our studies, we solely focused on FTV-induced illusions and presented a basic design space for FTV applications in virtual object interaction. 
However, additional user evaluations comparing these illusions with other kinesthetic haptic methods would shed light on the strengths and limitations of FTV-induced illusions, providing future researchers with a better design space for FTV usage. 
For instance, P14 for the rendering method evaluation commented on how the illusion was less apparent during the VR session with visual cues compared to the calibration stage. 
On the other hand, P5 for the VR experience evaluation felt an opposite experience, where the FTV illusion felt stronger with visual feedback. 
Exploring pseudo-haptic by manipulating control-to-display ratios with FTV-induced illusions would clarify how much visual information affects FTV perception.

While calibration typically took 5 to 10 minutes, there were participants who took up to 15~minutes. 
Needing to recognize the FTV illusion's unfamiliar sensation made the calibration tricky. The current calibration method is mostly trial and error, shifting the ERM motor location slightly and providing sample vibrations. 
For EMS, which also suffers from the same problem of long calibration time, there have been attempts to improve calibration time using EMG~\cite{knibbe2017automatic}. 
Since FTV induces Ia afferent discharges in the muscle~\cite{shinohara2005prolonged}, EMG also has a promising potential for decreasing calibration time for FTV.  

Different vibration parameters, such as vibration frequency, amplitude displacement, and duration, affect the amount of tendon vibration-induced illusion~\cite{taylor2017muscle}. 
To maintain a small form factor while delivering a strong enough vibration, we chose an ERM motor. However, for ERM motors, fixing the vibration frequency will fix the amplitude. 
In our work, this characteristic was insignificant because we focused on the effects of vibration duration. 
But for a more versatile use of FTV, additional studies on changes in frequency and amplitude would provide crucial information.
This may allow for different methods to control the amount of FTV illusion aside from duration, which would provide more adequate FTV illusions for virtual object interaction.

\section{Conclusion}
In this paper, we propose a novel method of providing kinesthetic haptic feedback for immersive virtual object interaction through FTV-induced illusions. 
Through 3~perception studies, we found that the illusion requires a minimum activation time of 0.75~s, does not significantly affect perceived voluntarily exerted finger force, and intensifies involuntary movement illusion over time.
Based on this, we developed 6~exemplary virtual object interaction scenarios ranging from UI components to 3D objects. In each scenario, the finger is subject to perceive either flexion or extension movement with our method. Our haptic rendering evaluation revealed that the illusion onset delay is negligible for long enough interactions, and tactile cues from longer vibrations are crucial for scenarios with constant force. 
Lastly, we compared FTV with simple vibration and no vibration in a VR evaluation study and found that FTV significantly improves VR experience, especially for body ownership. 
In particular, FTV was effective in providing users a sense of resistance on their finger, which the other two vibration settings could not. 
Using these results, we outlined a design guideline for FTV-induced illusions and additional applications. 
We hope these findings promote FTV's application in VR, providing users with high-quality, immersive VR experiences.

\begin{acks}
This research was supported by the MSIT~(Ministry of Science and ICT), Korea, under the ITRC~(Information Technology Research Center) support program~(IITP-2024-RS-2024-00436398) supervised by the IITP~(Institute for Information \& Communications Technology Planning \& Evaluation).
This work was supported by Electronics and Telecommunications Research Institute~(ETRI) grant funded by the Korean government~[26ZC1100, Development of Spatial Media Technology and Interaction Technology for Convergence of the Real and Virtual World].
\end{acks}

\bibliographystyle{ACM-Reference-Format}
\bibliography{newreferences}

\appendix
\section{ERM Start-Stop Delay}\label{appendix:erm}
Here, we provide the full measurements for our ERM delay. We sent target vibration durations and measured how much more the motor actually vibrated in milliseconds.

\begin{table}[h]
\caption{Delay in ms for each vibration durations.}
\label{table:erm delay}
\begin{tabular}{|c|c|c|c|c|c|}
\hline
Trials & 0.5s & 0.75s & 1.5s & 3.0s & 5.0s \\ \hline
1      & 53   & 58    & 84   & 45   & 65   \\ \hline
2      & 38   & 12    & 81   & 83   & 59   \\ \hline
3      & 40   & 39    & 53   & 26   & 61   \\ \hline
4      & 44   & 57    & 45   & 30   & 60   \\ \hline
5      & 36   & 62    & 60   & 26   & 44   \\ \hline
6      & 58   & 42    & 57   & 80   & 69   \\ \hline
7      & 43   & 54    & 52   & 14   & 47   \\ \hline
8      & 31   & 39    & 47   & 60   & 92   \\ \hline
9      & 30   & 17    & 63   & 54   & 42   \\ \hline
10     & 82   & 43    & 48   & 35   & 65   \\ \hline
\end{tabular}
\end{table}

\section{Perception Study 1: Effect on Perceived Pinching Force}\label{appendix:perception1} 
Here, we provide the full results for the perceived pinching force and the effect size. We compared the vibration settings for each averaged force in the 0.5~s time window.  

\begin{table}[t]
\caption{Mean, standard error, F-value, and p-value of each vibration setting.}
\label{table:pinching}
\begin{tabular}{|c|c|c|c|c|c|}
\hline
Time Window             & Setting   & Mean   & SE    & F-value               & p-value                \\ \hline
\multirow{4}{*}{0.5~s$\sim1.0$~s} & None      & 0.210  & 0.174 & \multirow{4}{*}{1.65} & \multirow{4}{*}{0.177} \\ \cline{2-4}
                     & Flexion   & -0.226 & 0.175 &                       &                        \\ \cline{2-4}
                     & Extension & 0.179  & 0.159 &                       &                        \\ \cline{2-4}
                     & Both      & 0.245  & 0.177 &                       &                        \\ \hline
\multirow{4}{*}{1.0~s$\sim1.5$~s} & None      & -0.158 & 0.254 & \multirow{4}{*}{0.68} & \multirow{4}{*}{0.565} \\ \cline{2-4}
                     & Flexion   & -0.448 & 0.242 &                       &                        \\ \cline{2-4}
                     & Extension & -0.069 & 0.244 &                       &                        \\ \cline{2-4}
                     & Both      & 0.017  & 0.240 &                       &                        \\ \hline
\multirow{4}{*}{1.5~s$\sim2.0$~s} & None      & -0.101 & 0.267 & \multirow{4}{*}{0.49} & \multirow{4}{*}{0.687} \\ \cline{2-4}
                     & Flexion   & -0.417 & 0.258 &                       &                        \\ \cline{2-4}
                     & Extension & 0.022  & 0.290 &                       &                        \\ \cline{2-4}
                     & Both      & -0.071 & 0.270 &                       &                        \\ \hline
\multirow{4}{*}{2.0~s$\sim2.5$~s} & None      & -0.048 & 0.272 & \multirow{4}{*}{0.35} & \multirow{4}{*}{0.790} \\ \cline{2-4}
                     & Flexion   & -0.281 & 0.269 &                       &                        \\ \cline{2-4}
                     & Extension & 0.040  & 0.277 &                       &                        \\ \cline{2-4}
                     & Both      & 0.082  & 0.282 &                       &                        \\ \hline
\multirow{4}{*}{2.5~s$\sim3.0$~s} & None      & -0.020 & 0.287 & \multirow{4}{*}{0.35} & \multirow{4}{*}{0.792} \\ \cline{2-4}
                     & Flexion   & -0.121 & 0.269 &                       &                        \\ \cline{2-4}
                     & Extension & 0.065  & 0.277 &                       &                        \\ \cline{2-4}
                     & Both      & 0.271  & 0.300 &                       &                        \\ \hline
\multirow{4}{*}{3.0~s$\sim3.5$~s} & None      & 0.008  & 0.301 & \multirow{4}{*}{0.49} & \multirow{4}{*}{0.686} \\ \cline{2-4}
                     & Flexion   & -0.106 & 0.264 &                       &                        \\ \cline{2-4}
                     & Extension & 0.057  & 0.299 &                       &                        \\ \cline{2-4}
                     & Both      & 0.369  & 0.291 &                       &                        \\ \hline
\multirow{4}{*}{3.5~s$\sim4.0$~s} & None      & 0.031  & 0.316 & \multirow{4}{*}{0.48} & \multirow{4}{*}{0.697} \\ \cline{2-4}
                     & Flexion   & -0.108 & 0.269 &                       &                        \\ \cline{2-4}
                     & Extension & 0.111  & 0.298 &                       &                        \\ \cline{2-4}
                     & Both      & 0.374  & 0.288 &                       &                        \\ \hline
\multirow{4}{*}{4.0~s$\sim4.5$~s} & None      & -0.052 & 0.308 & \multirow{4}{*}{0.67} & \multirow{4}{*}{0.571} \\ \cline{2-4}
                     & Flexion   & -0.112 & 0.265 &                       &                        \\ \cline{2-4}
                     & Extension & 0.175  & 0.305 &                       &                        \\ \cline{2-4}
                     & Both      & 0.421  & 0.304 &                       &                        \\ \hline
\multirow{4}{*}{4.5~s$\sim5.0$~s} & None      & 0.016  & 0.303 & \multirow{4}{*}{1.05} & \multirow{4}{*}{0.369} \\ \cline{2-4}
                     & Flexion   & -0.185 & 0.265 &                       &                        \\ \cline{2-4}
                     & Extension & 0.282  & 0.305 &                       &                        \\ \cline{2-4}
                     & Both      & 0.509  & 0.309 &                       &                        \\ \hline
\end{tabular}
\end{table}

\begin{table*}[h]
\caption{Cohen's f effect size for all comparison and time window. Blue number indicates small effect size. Yellow indicates medium effect size. Red indicates large effect size.}
\label{table:pinching effect}
\begin{tabular}{|c|c|c|c|c|c|c|c|}
\hline
Time Window           & \multicolumn{1}{l|}{Overall}  & \multicolumn{1}{l|}{None/Flexion} & \multicolumn{1}{l|}{None/Extension} & \multicolumn{1}{l|}{None/Both} & \multicolumn{1}{l|}{Flexion/Extension} & \multicolumn{1}{l|}{Flexion/Both} & \multicolumn{1}{l|}{Extention/Both} \\ \hline
0.5 s$\sim$1.0 s & {\color[HTML]{FF0000} 0.4611} & {\color[HTML]{FF9900} 0.3725}    & 0.0265                          & 0.0299                           & {\color[HTML]{FF9900} 0.346}    & {\color[HTML]{FF0000} 0.4024}    & 0.0564                          \\ \hline
1.0 s$\sim$1.5 s & {\color[HTML]{FF9900} 0.3535} & {\color[HTML]{FF9900} 0.2071}    & 0.0636                          & {\color[HTML]{0000FF} 0.125}     & {\color[HTML]{FF9900} 0.2707}   & {\color[HTML]{FF9900} 0.3321}    & 0.0614                          \\ \hline
1.5 s$\sim$2.0 s & {\color[HTML]{FF9900} 0.3173} & {\color[HTML]{FF9900} 0.2145}    & 0.0835                          & 0.0204                           & {\color[HTML]{FF9900} 0.298}    & {\color[HTML]{0000FF} 0.2349}    & 0.0631                          \\ \hline
2.0 s$\sim$2.5 s & {\color[HTML]{FF9900} 0.2678} & {\color[HTML]{0000FF} 0.1571}    & 0.0593                          & 0.0876                           & {\color[HTML]{0000FF} 0.2164}   & {\color[HTML]{0000FF} 0.2447}    & 0.0283                          \\ \hline
2.5 s$\sim$3.0 s & {\color[HTML]{FF9900} 0.271}  & 0.0671                           & 0.0565                          & {\color[HTML]{0000FF} 0.1933}    & {\color[HTML]{0000FF} 0.1236}   & {\color[HTML]{FF9900} 0.2604}    & {\color[HTML]{0000FF} 0.1368}   \\ \hline
3.0 s$\sim$3.5 s & {\color[HTML]{FF9900} 0.3274} & 0.075                            & 0.0322                          & {\color[HTML]{0000FF} 0.2375}    & {\color[HTML]{0000FF} 0.1072}   & {\color[HTML]{FF9900} 0.3125}    & {\color[HTML]{0000FF} 0.2053}   \\ \hline
3.5 s$\sim$4.0 s & {\color[HTML]{FF9900} 0.3244} & 0.0908                           & 0.0523                          & {\color[HTML]{0000FF} 0.2241}    & {\color[HTML]{0000FF} 0.1431}   & {\color[HTML]{FF9900} 0.315}     & {\color[HTML]{0000FF} 0.1719}   \\ \hline
4.0 s$\sim$4.5 s & {\color[HTML]{FF9900} 0.3864} & 0.039                            & {\color[HTML]{0000FF} 0.1476}   & {\color[HTML]{FF9900} 0.3076}    & {\color[HTML]{0000FF} 0.1867}   & {\color[HTML]{FF9900} 0.3467}    & {\color[HTML]{0000FF} 0.16}     \\ \hline
4.5 s$\sim$5.0 s & {\color[HTML]{FF0000} 0.4835} & {\color[HTML]{0000FF} 0.1307}    & {\color[HTML]{0000FF} 0.173}    & {\color[HTML]{FF9900} 0.3206}    & {\color[HTML]{FF9900} 0.3037}   & {\color[HTML]{FF0000} 0.4514}    & {\color[HTML]{0000FF} 0.1476}   \\ \hline
\end{tabular}
\end{table*}

\section{Rendering Method Evaluation: Questionnaires}\label{appendix:questionnaires}
Here, we provide the exact wording used in the questionnaires during the rendering method evaluation.
For body ownership, we used the same wordings for all scenarios.

\pagebreak

\textbf{Body Ownership}
\begin{itemize}
    \item \textbf{Q1}: I felt as if the virtual hand was my hand.
    \item \textbf{Q2}: It felt as if the virtual hand I saw was someone else's.
    \item \textbf{Q3}: It seemed as if I might have more than one hand.
\end{itemize}

\subsection{\textit{Button}}
\textbf{Tactile Sensation}
\begin{itemize}
\item \textbf{Q10}: It seemed as if I felt the touch of the button in the location where I saw the virtual hand touched.
\item \textbf{Q11}: It seemed as if the touch I felt was located somewhere between my physical hand and the virtual hand.
\item \textbf{Q12}: It seemed as if the touch I felt was caused by the button touching the virtual hand.
\item \textbf{Q13}: It seemed as if my hand was touching the button.
\end{itemize}

\subsection{\textit{Scrollbar}}
\textbf{Tactile Sensation}
\begin{itemize}
\item \textbf{Q10}: It seemed as if I felt the touch of the scrollbar in the location where I saw the virtual hand touched.
\item \textbf{Q11}: It seemed as if the touch I felt was located somewhere between my physical hand and the virtual hand.
\item \textbf{Q12}: It seemed as if the touch I felt was caused by the scrollbar touching the virtual hand.
\item \textbf{Q13}: It seemed as if my hand was touching the scrollbar.
\end{itemize}

\subsection{\textit{Rectangle}}
\textbf{Tactile Sensation}
\begin{itemize}
\item \textbf{Q10}: It seemed as if I felt the touch of the rectangle in the location where I saw the virtual hand touched.
\item \textbf{Q11}:v It seemed as if the touch I felt was located somewhere between my physical hand and the virtual hand.
\item \textbf{Q12}: It seemed as if the touch I felt was caused by the rectangle touching the virtual hand.
\item \textbf{Q13}: It seemed as if my hand was touching the rectangle.
\end{itemize}

\subsection{\textit{Train}}
\textbf{Tactile Sensation}
\begin{itemize}
\item \textbf{Q10}: It seemed as if I felt the touch of the train in the location where I saw the virtual hand touched.
\item \textbf{Q11}: It seemed as if the touch I felt was located somewhere between my physical hand and the virtual hand.
\item \textbf{Q12}: It seemed as if the touch I felt was caused by the train touching the virtual hand.
\item \textbf{Q13}: It seemed as if my hand was touching the train.
\end{itemize}
\textbf{3D Object Questionnaires}
\begin{itemize}
\item \textbf{Realism}: It felt as if I was pushing a real train.
\item \textbf{Immersion}: The sensation helped me become more immersed when pushing the train.
\item \textbf{Enjoyment}: The sensation helped me enjoy pushing the train.
\end{itemize}

\subsection{\textit{Ball}}
\textbf{Tactile Sensation}
\begin{itemize}
\item \textbf{Q10}: It seemed as if I felt the touch of the ball in the location where I saw the virtual hand touched.
\item \textbf{Q11}: It seemed as if the touch I felt was located somewhere between my physical hand and the virtual hand.
\item \textbf{Q12}: It seemed as if the touch I felt was caused by the ball touching the virtual hand.
\item \textbf{Q13}: It seemed as if my hand was touching the ball.
\end{itemize}
\textbf{3D Object Questionnaires}
\begin{itemize}
\item \textbf{Realism}: It felt as if I was holding a real ball.
\item \textbf{Immersion}: The sensation helped me become more immersed when holding the ball.
\item \textbf{Enjoyment}: The sensation helped me enjoy holding the ball.
\end{itemize}

\subsection{\textit{Kettlebell}}
\textbf{Tactile Sensation}
\begin{itemize}
\item \textbf{Q10}: It seemed as if I felt the touch of the kettlebell in the location where I saw the virtual hand touched.
\item \textbf{Q11}: It seemed as if the touch I felt was located somewhere between my physical hand and the virtual hand.
\item \textbf{Q12}: It seemed as if the touch I felt was caused by the kettlebell touching the virtual hand.
\item \textbf{Q13}: It seemed as if my hand was touching the kettlebell.

\end{itemize}
\textbf{3D Object Questionnaires}
\begin{itemize}
\item \textbf{Realism}: It felt as if I was pushing a real kettlebell.
\item \textbf{Immersion}: The sensation helped me become more immersed when lifting the kettlebell.
\item \textbf{Enjoyment}: The sensation helped me enjoy lifting the kettlebell.
\end{itemize}

\end{document}